\documentclass[lettersize,journal]{IEEEtran}
\usepackage{amsmath,amsfonts}
\usepackage{algorithmic}
\usepackage{algorithm}
\usepackage{array}
\usepackage[caption=false,font=normalsize,labelfont=sf,textfont=sf]{subfig}
\usepackage{textcomp}
\usepackage{stfloats}
\usepackage{verbatim}
\usepackage{graphicx}
\usepackage{cite}
\usepackage{hyperref}

\usepackage{pifont} 
\usepackage{cleveref}
\usepackage{xcolor}
\usepackage{colortbl}
\usepackage[most]{tcolorbox}
\usepackage{listings}
\usepackage{multirow} 
\usepackage{float} 
\usepackage{tikz}
\usepackage{booktabs} 
\usepackage{xspace}
\usepackage{enumitem}
\usepackage{tikz}

\definecolor{gptorange}{HTML}{FFC5C5}
\definecolor{deepgreen}{HTML}{CCFFCC}
\definecolor{OliveGreen}{rgb}{0,0.6,0}

\newcommand{\ourtool}[1]{\texttt{PoCSmith\xspace}}
\newcommand{\numreport}[1]{48\xspace}
\newcommand{\numconfirm}[1]{8\xspace}

\begin{document}
\title{A Systematic Study on Generating Web Vulnerability Proof-of-Concepts Using Large Language Models}

\author{~\IEEEmembership{Mengyao Zhao, Kaixuan Li, Lyuye Zhang, Wenjing Dang, Chenggong Ding, Sen Chen, and Zheli Liu
\thanks{Mengyao Zhao and Kaixuan Li contributed equally to this work.}
\thanks{Mengyao Zhao, Wenjing Dang, and Chenggong Ding are with the College of Intelligence and Computing, Tianjin University, Tianjin 300350, China (e-mail: mengyaozhao@tju.edu.cn; dangwenjing@tju.edu.cn; chenggong\_ding@tju.edu.cn).}
\thanks{Kaixuan Li and Lyuye Zhang are with the College of Computing and Data Science, Nanyang Technological University, Singapore (e-mail: kaixuan.li@ntu.edu.sg; zh0004ye@e.ntu.edu.sg).}
\thanks{Sen Chen (Corresponding author) and Zheli Liu are with the College of Cyber Science, Nankai University, Tianjin 300350, China (e-mail: senchen@nankai.edu.cn; liuzheli@nankai.edu.cn).}
}}

\markboth{Journal of \LaTeX\ Class Files,~Vol.~14, No.~8, August~2021}%
{Shell \MakeLowercase{\textit{et al.}}: A Sample Article Using IEEEtran.cls for IEEE Journals}

\IEEEpubid{0000--0000/00\$00.00~\copyright~2021 IEEE}

\maketitle

\begin{abstract}
Recent advances in Large Language Models (LLMs) have brought remarkable progress in code understanding and reasoning, creating new opportunities and raising new concerns for software security. Among many downstream tasks, generating Proof-of-Concept (PoC) exploits plays a central role in vulnerability reproduction, comprehension, and mitigation. While previous research has focused primarily on zero-day exploitation, the growing availability of rich public information accompanying disclosed CVEs leads to a natural question: can LLMs effectively use this information to automatically generate valid PoCs?

In this paper, we present the first empirical study of LLM-based PoC generation for web application vulnerabilities, focusing on the practical feasibility of leveraging publicly disclosed information. We evaluate GPT-4o and DeepSeek-R1 on 100 real-world and reproducible CVEs across three stages of vulnerability disclosure: (1) newly disclosed vulnerabilities with only descriptions, (2) 1-day vulnerabilities with patches, and (3) N-day vulnerabilities with full contextual code. Our results show that LLMs can automatically generate working PoCs in 8\%–34\% of cases using only public data, with DeepSeek-R1 consistently outperforming GPT-4o. Further analysis shows that supplementing code context improves success rates by 17\%-20\%, with function-level providing 9\%-13\% improvement than file-level ones. Further integrating adaptive reasoning strategies to prompt refinement significantly improves success rates to 68\%–72\%.
Our findings suggest that LLMs could reshape vulnerability exploitation dynamics. By lowering the technical barrier, they expand the potential attacker base. At the same time, the ability of LLMs to generate PoCs enables defenders to automate downstream tasks such as patch validation, regression testing, and vulnerability triage.
To date, 23 newly generated PoCs have been accepted by NVD and Exploit DB. 

\end{abstract}


\section{Introduction}
\IEEEPARstart{W}{eb} applications constitute the front line of modern digital services and absorb a large share of cyberattacks. Most of these services rely on PHP, which powers about 74\% of all active websites~\cite{Historic69:online}. Public vulnerability databases therefore list thousands of web flaws each year, including cross-site scripting (XSS), SQL injection, and cross-site request forgery (CSRF). For defenders, a reproducible \emph{Proof-of-Concept} (PoC) is indispensable for validating vulnerability, enabling root-cause analysis, and supporting regression testing. However, empirical studies show that nearly half of disclosed Common Vulnerabilities and Exposures (CVEs) either ship without a working PoC or provide one that third parties cannot reproduce~\cite{mu2018understanding}. 
Such reproducibility gaps necessitate significant manual reverse engineering by security teams to restore exploitable conditions~\cite{luo2022tchecker}.

\begin{figure}[]
  \centering
  \includegraphics[width= 0.85\linewidth]{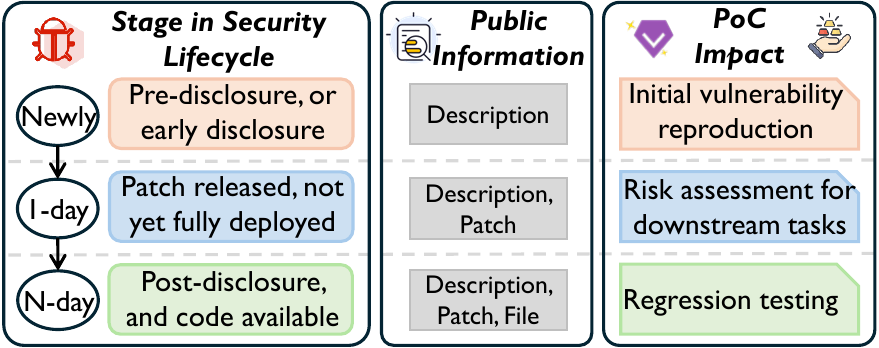}
  \caption{From disclosure to exploit: information availability and PoC impact.}
  \label{fig:Life_cycle.pdf}
\end{figure}

\IEEEpubidadjcol
\textbf{Limitations.}
\Cref{fig:Life_cycle.pdf} depicts the three typical stages of the vulnerability disclosure process, emphasizing the evolution of public information and the value of a PoC at each stage. Stage one involves a brief advisory marking the initial disclosure; stage two delivers a vendor patch within days, though adoption is only partial; and stage three, specific to open-source systems, follows broad patch deployment and shifts the analytical focus toward longitudinal exploitation dynamics and security research~\cite{project-zero}. 
Existing web PoC generation systems were mainly developed for zero-day research to reduce false positives in detection tools~\cite{kieyzun2009automatic,huang2013craxweb,alhuzali2016chainsaw, alhuzali2018navex,lee2020fuse,chen2023uradar, park2022fugio}, exhibiting three critical gaps for disclosed vulnerabilities: 
\textbf{\textit{1)}} Most require direct access to vulnerable code (stage three)~\cite{alhuzali2016chainsaw,chen2023uradar,park2022fugio}, few can use patch diffs (stage two)~\cite{shi2024recurscan}, and none operate when only description is available (stage one). Consequently, they cover only part of the disclosure timeline and leave much public data untapped. 
\textbf{\textit{2)}} These methods mainly depend on expert-crafted attack templates that fail to generalize across evolving vulnerability patterns~\cite{kieyzun2009automatic,huang2013craxweb,lee2020fuse, alhuzali2018navex}; 
\textbf{\textit{3)}} and their symbolic execution back-ends struggle with the scale and dynamic features of modern web stacks such as nested sanitization logic and framework-specific templating~\cite{huang2021ufuzzer, alhuzali2016chainsaw}.
These gaps make it difficult to generate reliable PoCs for disclosed vulnerabilities and highlight the need for a new approach that leverages the full spectrum of public artifacts.

{\textbf{Motivation.}
The limitations discussed above necessitate a new perspective on PoC construction within real-world disclosure workflows.
Recent LLMs have shown the ability to process heterogeneous inputs such as natural language advisories, patches, and raw source files, and then synthesize coherent executable code~\cite{kang2023large, zhang2024acfix, deng2024pentestgpt, zhou2024magneto}.
In principle, an LLM could act as an adaptive exploit engineer that refines its output as richer artifacts become available at successive disclosure stages, thus avoiding the need for expert-crafted grammars and the path explosion limits of symbolic execution.
If this capability proves reliable, the security implications are immediate.
Defenders could generate regression tests within minutes after a CVE description is published, whereas attackers could shorten the interval between disclosure and weaponization.
Hence, sound threat modeling requires a systematic evaluation of how effectively LLMs can transform progressively disclosed information into functional PoCs for web vulnerabilities.}

\textbf{Approach.}
To this end, we design the first systematic study that follows the real chronology of vulnerability disclosure in~\Cref{fig:Life_cycle.pdf}. Practice shows three canonical stages: a newly pre-disclosure CVE that offers only a brief description, a 1-day CVE for which a patch has been released, and an N-day CVE whose vulnerable source code is open-source and publicly available~\cite{ProjectZ60:online}. 
Using these public information, we conduct a progressive four-phase empirical evaluation of two representative LLMs, GPT-4o and DeepSeek-R1. The benchmark consists of 100 reproducible CVEs spanning 5 web vulnerability categories (CWE-78/79/89/352/434). Our investigation is guided by 4 research questions: (RQ1) How well can LLMs generate functional PoCs when limited to the public artifacts at each disclosure stage? (RQ2) What are the root causes of generation failures, and how can hierarchical task decomposition help to expose them? (RQ3) How does the granularity of added code context, such as file-level versus function-level, affect success rates? (RQ4) Can iterative feedback-driven adaptive prompt strategies, such as chain-of-thought reasoning and in-context learning, further improve effectiveness? 
These questions provide a structured perspective for assessing current capabilities and for guiding the security community on whether to integrate powerful LLMs into future PoC workflows or instead place constraints on their use.

\textbf{Findings.} 
Our findings illuminate both the promise and the peril of LLMs in automated vulnerability exploitation:  
\textbf{\textit{1)}} Even minimal public disclosure enables exploitation. When given only a description, LLMs produce valid PoCs for 8\% of the evaluated CVEs. 
LLMs exhibit an emerging capability to synthesize PoCs directly from public information, signaling a new paradigm in PoC generation. 
\textbf{\textit{2)}} {Further failure analysis uncovers critical context sensitivity and inherent capability boundaries: 86.4\% of failures stem from missing navigation context like file hierarchies, and LLM's native reasoning accuracy reaches only 65.1\% (GPT-4o).}
\textbf{\textit{3)}} Context supplementation demonstrates LLMs' environmental sensitivity: supplementing function-level context elevates success rates to 54\% (DeepSeek-R1) and 38\% (GPT-4o), validating contextual completeness as a critical success factor. 
\textbf{\textit{4)}} Feedback-driven adaptive prompt strategies including CoT and ICL could narrow the capability gap: GPT-4o reaches 68\% effectiveness, nearing DeepSeek-R1's 72\%, suggesting task-aware scaffolding can mitigate inherent limitations. 
14 generated PoCs have been accepted by the NVD~\cite{DatabaseNVD} and 9 by Exploit DB~\cite{ExploitD46:online}, confirming the practicality. 
These results underline two clear implications: First, the paradigm offers defenders new opportunities to accelerate vulnerability triage, streamline regression testing, and simulate exploit behavior at scale. 
Second, these same capabilities may drastically lower the barrier to weaponization for attackers, especially within the critical window between disclosure and patch deployment. The security community may need to reassess how disclosure policies strike the balance between transparency and the unprecedented generative power now afforded by LLMs.

In summary, we make the following four contributions: 
\begin{itemize}[leftmargin=5pt]
    \item \textbf{Originality:} We conduct the first empirical study to systematically evaluate the PoC generation capabilities and limitations of LLMs across three security disclosure scenarios, spanning 100 reproducible real-world web vulnerabilities (5 types) with nearly 3,000 experimental trials. 

    \item \textbf{Evaluation:} We conduct an in-depth evaluation by a 4-phase methodology: basic effectiveness, failure cause analysis via sub-task decompositions, impact quantification of context supplementation, and adaptive reasoning optimization. {It requires 600+ man-hours to build reproduction environments and 121+ man-hours to validate PoCs.}

    \item \textbf{Open science:} 
    We provide three key resources to advance research in LLM-driven PoC generation\footnote{We will release all the evaluation data upon the acceptance of this work.}: LLM-driven PoC generation prompts that combine CoT, ICL, and a real-time feedback mechanism to effectively unleash the potential of LLMs; a curated benchmark comprising 100 reproducible CVEs with ground-truth PoCs and structured contextual information; and all experimental results. 
    \item \textbf{Insights:} 
    {Our results uncover a new paradigm in LLM-driven PoC generation from public disclosures, compelling a careful re-examination of disclosure practices to balance transparency against generative capability. We further dissect this phenomenon and explore how contextual completeness and adaptive reasoning prompt strategies can significantly enhance their exploitation potential.}
\end{itemize}

\section{Background and Study Design}

\begin{figure*}[]
  \centering
  \includegraphics[width=0.9\linewidth]{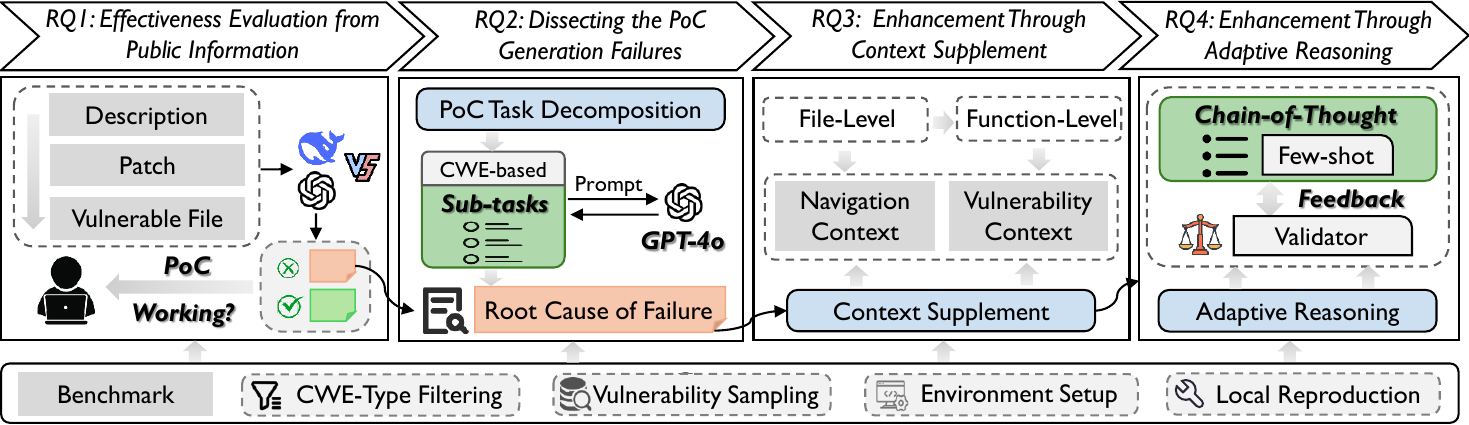}
  \caption{Overview of our study.} 
  \label{fig:overview}
\end{figure*}

\subsection{Web Applications and Vulnerabilities}
Web applications serve as the main vehicle for delivering information and services on the modern Internet. Due to their widespread adoption and the valuable data they store, they are often the preferred target of attackers~\cite{WhatareW7:online}. The representative web vulnerability types include: Cross-site Scripting (CWE-79), SQL Injection (CWE-89), Cross-Site Request Forgery (CWE-352), OS Command Injection (CWE-78), and Unrestricted File Upload (CWE-434). 
They are widely researched in the cybersecurity field and consistently ranked highly in the latest ``CWE Top 25 Most Dangerous Software Weaknesses''~\cite{CWE2024C67:online}. 
A more detailed selection principle is provided in~\Cref{sec:benchmark}. 
Moreover, since CWE-78, CWE-79, and CWE-89 typically rely on user input without adequate sanitization before reaching security-sensitive operations (i.e., sinks), we collectively refer to them as taint-style vulnerabilities~\cite{luo2022tchecker}.

\subsection{Large Language Model (LLM)} 
LLMs have demonstrated strong capabilities in code understanding and have been widely applied in cybersecurity tasks such as penetration testing~\cite{deng2024pentestgpt}, vulnerability detection~\cite{ullah2024llms}, and vulnerability repair~\cite{zhang2024acfix}. Yet, their potential in generating PoCs for web vulnerabilities remains unexplored.
Therefore, our study aims to investigate the capabilities and limitations of LLMs to generate PoCs and provide valuable insights into this critical research area. 
To do this, we choose two representative LLMs (GPT-4o~\cite{hurst2024gpt} and DeepSeek-R1~\cite{guo2025deepseek}) in our study.
GPT-4o is the latest generic LLM, known for its code generation proficiency at the time of our study. 
DeepSeek-R1 produces long CoT steps to improve reasoning abilities, making it suitable for complex problems. While open-source alternatives such as Llamma3~\cite{grattafiori2024llama}, StarCoder~\cite{li2023starcoder}, and CodeLamma~\cite{roziere2023code} exist, they remain unselected since they underperform in security tasks (20\% lower F1~\cite{zhou2024large,fang2024large,alam2024ctibench}).

\subsection{Study Design}
As shown in~\Cref{fig:overview}, to assess whether LLMs can effectively understand and utilize publicly disclosed information aligned with realistic threat models, we conducted an in-depth and progressive exploration of their effectiveness and limitations in PoC generation through four interconnected RQs: 
(1) RQ1: Effectiveness Evaluation from
Public Information. We begin by evaluating baseline performance using basic prompts and public information (e.g., vulnerability descriptions, patch commits, and related vulnerable files) to gauge raw generation ability in three unique exploitation scenarios, as shown in~\Cref{sec:step1}. These inputs reflect realistic attacker knowledge at different disclosure stages, namely, newly disclosed, 1-day, and N-day vulnerabilities (~\Cref{fig:Life_cycle.pdf}). 
(2) RQ2: Dissecting the PoC Generation Failures. 
Based on RQ1's evaluation results, we further decompose the PoC generation process into specific sub-tasks to systematically identify the points at which LLMs fail, as shown in~\Cref{sec:step2}. The results informed the subsequent two optimization stages.
(3) RQ3: Enhancement Through Context Supplement. 
Then, for cases where context deficiency causes failure, we supplement missing vulnerability context and navigation context at two granularities, i.e., file-level and function-level. This allows us to further evaluate how context richness affects exploitability, as shown in~\Cref{sec:step3}.
(4) RQ4: Enhancement Through Adaptive Reasoning. 
We further combine chain-of-thought (CoT), in-context learning (ICL) and real-time feedback mechanism to unlock the potential of LLMs in PoC generation, as shown in~\Cref{sec:step4}. 
This progressive design allows us to systematically explore how different input factors influence the effectiveness of LLM-driven PoC generation.

\begin{figure*}[th]
  \centering
  \includegraphics[width=0.9\linewidth]{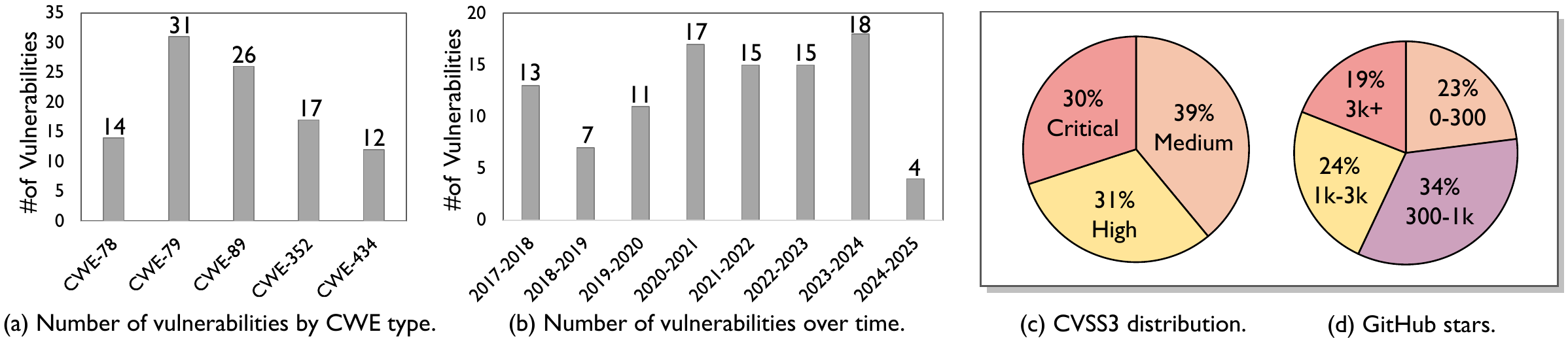}
  \caption{Benchmark of our study.}
  \label{fig:benchmark}
\end{figure*}

\section{Benchmark of Real-World Vulnerabilities} 
In this section, we construct a reproducible vulnerability benchmark as the basis for subsequent evaluations of LLMs to generate PoCs. In the process of constructing the dataset, we emphasize its representativeness and practicality.

\subsection{Vulnerability Type Selection}\label{sec:benchmark}
 
Given the impracticality of discussing each vulnerability type in detail, we try to select representative vulnerabilities in web applications. Specifically, we first conducted a systematic literature review (SLR) to collect web PoC generation papers to identify commonly studied vulnerability types~\cite{martin2008automatic,kieyzun2009automatic, bisht2011waptec,alhuzali2016chainsaw,pellegrino2017deemon,jan2017automatic,lee2020fuse,park2022fugio,zhao2023remote,cao2023oddfuzz,chenEfficientDetectionJava2024} published over the past two decades on TIFS, CCS, USENIX, NDSS, S\&P, ICSE, ISSTA, etc. 
We then cross-referenced these findings with the ``CWE Top 25 Most Dangerous Software Weaknesses''~\cite{CWE2024C67:online} list to select the top 5 most dangerous vulnerabilities. As shown in~\Cref{fig:benchmark}a, we ultimately determined these vulnerability types: CWE-79 (\textit{Cross-site Scripting, ranked 1st}), CWE-89 (\textit{SQL Injection, ranked 3rd}), CWE-352 (\textit{Cross-Site Request Forgery, ranked 4th}), CWE-78 (\textit{OS Command Injection, ranked 7th}) and CWE-434 (\textit{Unrestricted File Upload, ranked 10th}). These types are commonly researched in the cybersecurity field and highly ranked in the list, underscoring their prevalence and severity.  

\subsection{Vulnerability Sampling and Reproduction}
To ensure dataset timeliness, we first crawled all CVE entries from the National Vulnerability Database (NVD)~\cite{DatabaseNVD} between 2017 and 2024 that matched the selected vulnerability types. 
We then filtered vulnerabilities with a CVSS3~\cite{CommonVu49:online} score above 4.0 (at least "Medium" severity) to focus on those vulnerabilities with practical security threat. Subsequently, to ensure real-world reproducibility while avoiding manual bias, we randomly selected one vulnerability from each of the five types in every round and attempted to reproduce it. If the vulnerability was open-source and successfully reproduced, it was included in the dataset; otherwise, it was discarded. This process continued until we collected 100 vulnerabilities. In total, we conducted 35 rounds of sampling, including 31 CWE-79, 26 CWE-89, 17 CWE-352, 14 CWE-78, and 12 CWE-434 vulnerabilities, as shown in~\Cref{fig:benchmark}a. 

\noindent \textbf{Ethical considerations.} Ensuring each selected vulnerability was genuinely reproduced was critical to isolate LLM failures from improper environment setup. 
To prevent any unintended harm, we established secure and reliable local reproduction environments in controlled modes following established ethical principles~\cite{CodeofEt67:online}. However, the biggest challenge is that reproducing a vulnerability requires almost exclusively \textit{manual efforts}, and requires the reproducer to have highly specialized knowledge and skill sets~\cite{mu2018understanding}. 
For each vulnerability, we meticulously replicated the conditions required for reproduction, involving the deployment of the exact software version, installation of dependencies, and configuration of other services. The entire reproduction process required over 600+ man-hours, resulting in a fully validated and reproducible vulnerability benchmark. Additionally, we clarify that our study focuses on benign PoC generation for vulnerability validation using LLMs, rather than generating malicious exploits for attack.

\subsection{Preliminary Analysis}
We conduct a preliminary statistical analysis of the benchmark. As shown in~\Cref{fig:benchmark}, our dataset is highly representative in terms of vulnerability types, time distribution, severity, and real-world impact, providing a robust foundation for evaluating LLMs' capabilities. Additionally, as the dataset grew, the cost of environment reconstruction and PoC validation increased significantly, and the high intra-CWE redundancy in exploitation patterns yields limited returns. Therefore, considering the reproduction cost, the representativeness of vulnerability categories, and the diversity of exploitation scenarios, we clarify that our benchmark strikes a reasonable balance between experimental depth and resource feasibility.

\begin{table*}[]
\caption{Direct PoC generation results under varying inputs.}  
\label{tab:rq1}
\centering
\scalebox{1}{
\begin{tabular}{cc|c|ccc|cccccc}
\hline
\multicolumn{2}{c|}{\textbf{Vulnerability}}                                            &                                 & \multicolumn{3}{c|}{\textbf{PoC \textit{Rate}\textit{\textsubscript{success}}}}                               & \multicolumn{6}{c}{\textbf{PoC Format \textit{Rate}\textit{\textsubscript{distribution}}}}                                                                                                                                                                                                                     \\ \cline{1-2} \cline{4-12} 
\multicolumn{1}{c|}{\textbf{Type}}                    & \textbf{\#Vulns}               & \multirow{-2}{*}{\textbf{LLMs}} & \textbf{$\mathcal{S}_1$} & \textbf{$\mathcal{S}_2$} & \multicolumn{1}{c|}{\textbf{$\mathcal{S}_3$}}    & \textbf{HTML}                           & \textbf{Python}                & \textbf{\begin{tabular}[c]{@{}c@{}}Shell\\ Command\end{tabular}} & \textbf{Text}                  & \textbf{\begin{tabular}[c]{@{}c@{}}Proxy-style\\ Request\end{tabular}} & \textbf{PHP}   \\ \hline
\multicolumn{1}{c|}{}                                 &                                & GPT-4o                          & 7.1\%           & 14.3\%          & \cellcolor[HTML]{FFC5C5}21.4\%          & 0.8\%                                   & \cellcolor[HTML]{FFC5C5}45.2\% & 28.6\%                                                           & 2.4\%                          & -                                                                      & 23.0\%         \\
\multicolumn{1}{c|}{\multirow{-2}{*}{CWE-78}}         & \multirow{-2}{*}{14}           & DeepSeek-R1                     & 7.1\%           & 21.4\%          & \cellcolor[HTML]{CCFFCC}35.7\%          & -                                       & 26.4\%                         & 7.1\%                                                            & 2.9\%                          & \cellcolor[HTML]{CCFFCC}62.9\%                                         & 0.7\%          \\ \hline
\multicolumn{1}{c|}{}                                 &                                & GPT-4o                          & 9.7\%           & 16.1\%          & \cellcolor[HTML]{FFC5C5}22.6\%          & \cellcolor[HTML]{FFC5C5}85.7\%          & -                              & -                                                                & 11.4\%                         & 2.9\%                                                                  & -              \\
\multicolumn{1}{c|}{\multirow{-2}{*}{CWE-79}}         & \multirow{-2}{*}{31}           & DeepSeek-R1                     & 9.7\%           & 29.0\%          & \cellcolor[HTML]{CCFFCC}38.7\%          & 14.4\%                                  & -                              & -                                                                & \cellcolor[HTML]{CCFFCC}68.0\% & 17.6\%                                                                 & -              \\ \hline
\multicolumn{1}{c|}{}                                 &                                & GPT-4o                          & 7.7\%           & 11.5\%          & \cellcolor[HTML]{FFC5C5}23.1\%          & 3.0\%                                   & \cellcolor[HTML]{FFC5C5}53.4\% & 14.1\%                                                           & 12.4\%                         & 17.1\%                                                                 & -              \\
\multicolumn{1}{c|}{\multirow{-2}{*}{CWE-89}}         & \multirow{-2}{*}{26}           & DeepSeek-R1                     & 7.7\%           & 11.5\%          & \cellcolor[HTML]{CCFFCC}34.6\%          & 0.4\%                                   & 6.3\%                          & -                                                                & 2.5\%                          & \cellcolor[HTML]{CCFFCC}90.8\%                                         & -              \\ \hline
\multicolumn{1}{c|}{}                                 &                                & GPT-4o                          & 5.9\%           & 11.8\%          & \cellcolor[HTML]{FFC5C5}17.6\%          & \cellcolor[HTML]{FFC5C5}100.0\%         & -                              & -                                                                & -                              & -                                                                      & -              \\
\multicolumn{1}{c|}{\multirow{-2}{*}{CWE-352}}        & \multirow{-2}{*}{17}           & DeepSeek-R1                     & 5.9\%           & 23.5\%          & \cellcolor[HTML]{CCFFCC}29.4\%          & \cellcolor[HTML]{CCFFCC}100.0\%         & -                              & -                                                                & -                              & -                                                                      & -              \\ \hline
\multicolumn{1}{c|}{}                                 &                                & GPT-4o                          & 8.3\%           & 8.3\%           & \cellcolor[HTML]{FFC5C5}16.7\%          & 1.9\%                                   & \cellcolor[HTML]{FFC5C5}71.3\% & 8.3\%                                                            & 7.4\%                          & 2.8\%                                                                  & 8.3\%          \\
\multicolumn{1}{c|}{\multirow{-2}{*}{CWE-434}}        & \multirow{-2}{*}{12}           & DeepSeek-R1                     & 8.3\%           & 8.3\%           & \cellcolor[HTML]{CCFFCC}25.0\%          & -                                       & \cellcolor[HTML]{CCFFCC}68.6\% & 2.5\%                                                            & 0.8\%                          & 28.1\%                                                                 & -              \\ \hline
\multicolumn{1}{c|}{}                                 &                                & GPT-4o                          & \textbf{8.0\%}  & \textbf{13.0\%} & \cellcolor[HTML]{FFC5C5}\textbf{21.0\%} & \cellcolor[HTML]{FFC5C5}\textbf{44.6\%} & \textbf{28.8\%}                & \textbf{8.7\%}                                                   & \textbf{8.0\%}                 & \textbf{5.7\%}                                                         & \textbf{4.2\%} \\
\multicolumn{1}{c|}{\multirow{-2}{*}{\textbf{Total}}} & \multirow{-2}{*}{\textbf{100}} & DeepSeek-R1                     & \textbf{8.0\%}  & \textbf{20.0\%} & \cellcolor[HTML]{CCFFCC}\textbf{34.0\%} & \textbf{20.8\%}                         & \textbf{14.3\%}                & \textbf{1.4\%}                                                   & \textbf{22.1\%}                & \cellcolor[HTML]{CCFFCC}\textbf{41.3\%}                                & \textbf{0.1\%} \\ \hline
\end{tabular}
}
\end{table*}

\section{Empirical Study}
We first evaluate the baseline effectiveness of LLMs under varying levels of public information disclosure, then perform failure analysis through task decomposition, and finally supplement missing context and design adaptive prompts to improve PoC generation effectiveness, systematically studying the capabilities and potential of LLMs in PoC generation. 

\subsection{RQ1: Effectiveness Evaluating from Public Information}\label{sec:step1} 
In this section, we explore the original capabilities of LLMs to autonomously generate PoCs from public disclosed information with base prompt. To emulate realistic threat models, we design three unique input scenarios that progressively reflect different stages of vulnerability disclosure.

\subsubsection{Setup}
The vulnerability disclosure process typically reveals three types of public information: vulnerability descriptions, patch commits showing security fixes, and affected source code files. These elements, accessible via public databases (e.g., NVD~\cite{DatabaseNVD}) and code repositories, provide the essential basis for both human analysts and automated systems to understand and exploit vulnerabilities~\cite{you2017semfuzz}. We formalize the \textit{Public Information} for a vulnerability $v$ as a triplet $\mathcal{P}(v) = \langle D, P, F \rangle$, where $D$ denotes the natural-language description of the vulnerability~\cite{you2017semfuzz}, $P$ denotes the code differences between patched and vulnerable versions~\cite{brumley2008automatic}, and $F$ denotes the complete source code of files modified by $P$~\cite{alhuzali2018navex}.

\begin{figure}[]
  \centering
  \includegraphics[width=0.9\linewidth]{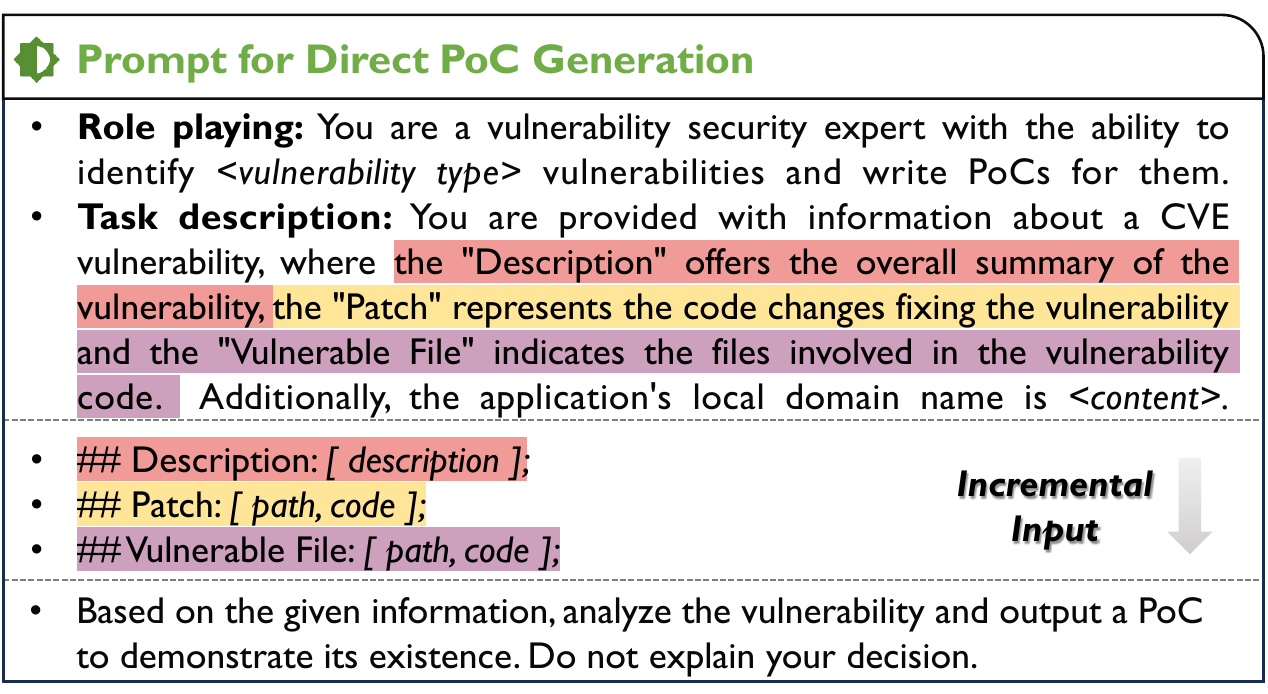}
  \caption{Basic prompt template for PoC generation.}
  \label{fig:RQ1_Prompt}
\end{figure}

{To explore the basic ability of LLMs to directly generate PoCs, we set up a base prompt and input public information in three combinations, reflecting realistic disclosure scenarios. As shown in~\Cref{fig:Life_cycle.pdf}, we define three input scenarios $\mathcal{S} = \{\mathcal{S}_1, \mathcal{S}_2, \mathcal{S}_3\}$ aligns with realistic threat for each vulnerability $v$ based on its public information $\mathcal{P}(v)$, where:}
\begin{itemize}[leftmargin=5pt]
    \item $\mathcal{S}_1 = \{D\}$: Using only the description, reflecting minimal disclosure but with potentially semantic guidance, corresponding to the newly disclosed vulnerability.
    \item $\mathcal{S}_2 = \{D, P\}$: Combining the description and patch, supporting semantic understanding and vulnerability localization, corresponding to the 1-day vulnerability.
    \item $\mathcal{S}_3 = \{D, P, F\}$: Incorporating all directly accessible public information without any additional technical analysis or supplement, corresponding to the N-day open vulnerability.
\end{itemize}
This design encompasses the entire disclosure timeline, allowing a nuanced evaluation of LLM's effectiveness under real-world threat scenarios.

As shown in ~\Cref{fig:RQ1_Prompt}, our prompt includes two parts: (1) the natural language part that explains the task to LLM and (2) the information input part that contains different combinations of public information. For the natural language part, we embed: (1) a role-playing instruction to inspire LLMs' PoC generation capability of different vulnerability types, assigned in the ``system'' prompt; and (2) a task-description instruction to explain the different information inputs and task outputs. We also need to provide the application's local domain name (e.g., http://phpmyadmin), which refers to the domain used to access the application in a local testing environment, essential for accurate replication. For the information input part, it corresponds to the three disclosure scenarios outlined earlier. Additionally, the task description instruction changes as the input scenario evolves.

Given the inherent instability of LLM outputs under identical prompts~\cite{ullah2024llms}, we adopted a systematic evaluation approach. The temperature parameter was set to 0 to reduce generation randomness, with each vulnerability executed in three independent trials. 
A vulnerability was considered successfully reproduced only if all three generations produced functional PoCs, a conservative criterion designed to filter out unstable one-off generations and reflect a consistent understanding of vulnerability of the models~\cite{feng2024prompting}. Lower thresholds may inflate the success rate, but risk overstating the models’ true capability.
To validate effectiveness, we manually executed the generated PoCs in controlled local environments.

We use \textit{$\text{Rate}_{\text{success}}$} and \textit{$\text{Rate}_{\text{distribution}}$} to represent the PoC generation success rate and the distribution of PoC formats, respectively. The quantification metrics are defined as: 
$\textit{Rate}_{\textit{success}} = \frac{N_{\text{success}}}{N_{\text{vulns}}}$ and
$\textit{Rate}_{\textit{distribution}} = \frac{N_{\text{format}}}{N_{\text{PoC}}}$, where $N_{\text{success}}$ denotes total vulnerabilities with successful PoCs, and $N_{\text{vulns}}$ denotes total vulnerabilities in that vulnerability type. $N_{\text{format}}$ denotes total PoCs with a specific format, and $N_{\text{PoC}}$ denotes all PoCs generated across all three input scenarios and three repeated trials in that vulnerability type.

\subsubsection{Results} 

As input richness progresses along the three disclosure stages, PoC generation effectiveness of both models steadily improve. 
As shown in~\Cref{tab:rq1}, when only vulnerability description $D$ (stage one) is provided, GPT-4o and DeepSeek-R1 succeed on 8.0\% of cases. We analyzed these description styles and found that valid PoCs were generated when the description included either explicit exploitation details (3 cases) or clearly root causes, such as vulnerable component and parameters, as long as no additional mitigation in the code (5 cases), as shown in~\Cref{fig:RQ1_description.pdf}. 
It shows that LLMs can infer exploitation semantics purely from natural-language artifacts.
With the inclusion of patch commit $P$ (stage two), success increases to 13.0\% and 20.0\%, as $P$ often reveals the root cause~\cite{shi2024recurscan}. 
Finally, when an additional vulnerable file $F$ is provided (stage three), LLMs gain more code semantics (e.g., data and control flow constraints), leading to further improvements: 21.0\% for GPT-4o and 34.0\% for DeepSeek-R1.
This trend holds across different vulnerability types.

\begin{tcolorbox}[size=title,opacityfill=0.1,breakable]
\noindent \textit{{\textbf{Finding 1:}}} {\textit{PoC generation effectiveness steadily improves across three disclosure stages, with 8\% in $\mathcal{S}1$, 12–20\% in $\mathcal{S}2$, and 21–34\% in $\mathcal{S}3$, reflecting LLMs’ capability to extract and operationalize latent vulnerability semantics from natural language and code artifacts in the full disclosure timeline. LLMs exhibit a new paradigm in PoC generation across vulnerability disclosure lifecycle.}}
\end{tcolorbox} 

{\Cref{tab:rq1} shows that LLMs can automatically generate functional PoCs for 8.0\%–34.0\% of vulnerabilities using only public information disclosure, revealing an emerging capability that may reshape the PoC generation landscape.
Analysis across CWE-IDs indicates that both models can generalize across vulnerability types from their pre-trained knowledge, performing slightly better on taint-style flaws (i.e., CWE-78, CWE-79, and CWE-89) than on logic-oriented ones (CWE-352 and CWE-434). This gap is more pronounced for DeepSeek-R1, which achieves 38.7\% on CWE-79 versus 25.0\% on CWE-434, whereas GPT-4o shows more uniform but overall lower effectiveness. 
These observations highlight two key insights. First, LLMs can autonomously reconstruct exploitation logic guided solely by public information, revealing that seemingly low-risk and seemingly benign disclosure artifacts may carry latent security implications. Second, the overall low success rates expose fundamental limitations in LLMs’ original abilities, motivating a deeper investigation into the reasons behind generation failures.}

\begin{tcolorbox}[size=title,opacityfill=0.1,breakable]
\noindent \textit{\textbf{Finding 2:}} {
\textit{Using only public disclosures, LLMs synthesize functional PoCs in 8\%–34\% of cases. Both LLMs perform slightly better on taint-style vulnerabilities, with DeepSeek-R1 outperforming GPT-4o on our benchmark. This emerging capability compels disclosure policies to strike a careful balance between transparency and the novel generative power afforded by LLMs.}
}
\end{tcolorbox}

We observe that LLMs exhibit certain limitations in PoC generation under public information, particularly in crafting attack payloads for taint-style vulnerabilities  (i.e., CWE-78/79/89). 
The attack payload is a critical component in constructing PoCs for taint-style vulnerabilities, directly determining whether the vulnerability can be successfully triggered.

\begin{figure}%[th]
  \centering
  \includegraphics[width=0.9\linewidth]{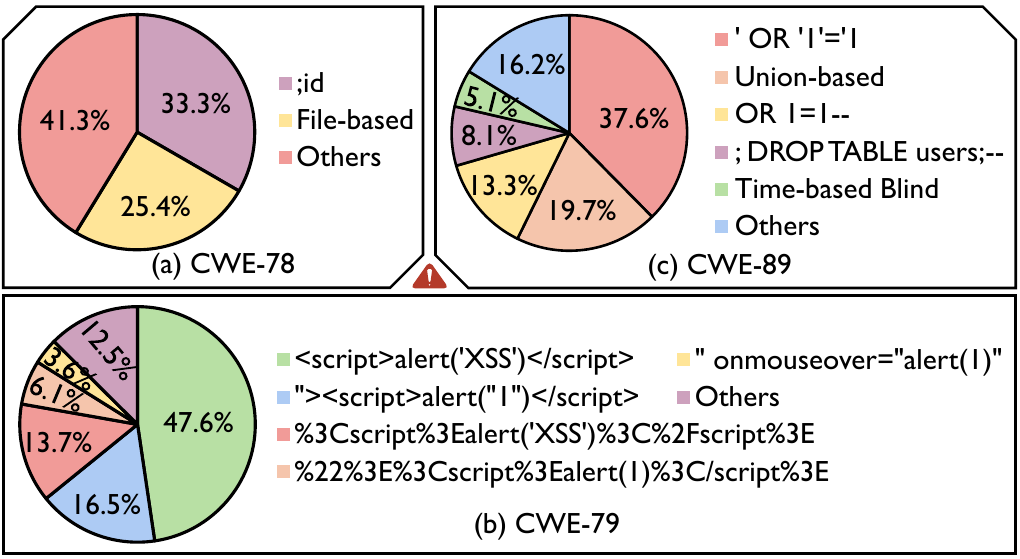}
  \caption{Distribution of taint-style vulnerability attack payloads generated by GPT-4o.} 
  \label{fig:RQ1_payload}
\end{figure}

We analyzed the distribution of attack payloads generated by two LLMs, as shown in~\Cref{fig:RQ1_payload} and ~\Cref{fig:RQ1_Payload_R1.pdf} in Appendix~\ref{statistics}, and observed that GPT-4o heavily relies on basic patterns, while DeepSeek-R1 demonstrates greater diversity and complexity. 
For example, for CWE-79, 47.6\% of attack payloads follow the basic \texttt{<script>alert('XSS')</script>} pattern and 16.5\% use \texttt{"><script>alert("1")</script>} by GPT-4o. 
However, this distribution seems to differ from real-world patterns. 
As demonstrated in XSS vulnerability submissions to the XSSED platform~\cite{buyukkayhan2020s}, advanced context-aware payload constructions (e.g., HTML context escapes requiring precise delimiter breaking) dominate practical exploitation scenarios, while elementary payload templates remain rare. 
In contrast, DeepSeek-R1 generates advanced context-aware payloads at least 33.4\%, reflecting its superior context awareness and reasoning ability. 
Such disparity suggests that, when contextual signals are limited (only public information available) and basic prompts are used, LLMs may struggle to generate effective payloads against modern sanitization defenses.

\begin{tcolorbox}[size=title,opacityfill=0.1,breakable]
\noindent \textit{{\textbf{Finding 3:}}} {\textit{
When synthesizing attack payloads from public information, DeepSeek-R1 generates more diverse payloads, while GPT-4o produces simplistic patterns, deviating from realistic exploitation. Nevertheless, LLMs still exhibit the potential to transcend rigid, expert-crafted templates and adapt to evolving vulnerability patterns.
}
}
\end{tcolorbox}

We further observe that LLMs exhibit format tendencies when generating PoCs. As shown in~\Cref{tab:rq1}, they demonstrate consistent formats for CWE-352 and CWE-434, predominantly using HTML and Python formats, respectively. For other three vulnerability types, the output formats show differences. 
Overall, GPT-4o tends to generate directly executable code-style PoCs, with HTML (44.6\%) and Python (28.8\%) being the dominant formats. In contrast, DeepSeek-R1 focuses more on interaction-based formats, such as proxy-style HTTP requests (41.3\%) and plain-text constructions (22.1\%), which more closely resemble the behavior of real-world manual penetration testing, as illustrated in~\Cref{fig:RQ1_PoC_Format.pdf}. 
This difference may stem from fundamental distinctions in training objectives and reasoning mechanisms. DeepSeek-R1 emphasizes a structured reasoning process, incorporating multi-stage reinforcement learning to align the model with human preferences~\cite{guo2025deepseek}. 
In contrast, GPT-4o develops robust reasoning skills by exposure to structured code logic and problem-solving processes, focusing on solving problems efficiently~\cite{hurst2024gpt}.

\begin{tcolorbox}[size=title,opacityfill=0.1,breakable]
\noindent \textit{\textbf{Finding 4:}} {\textit{LLMs have shown tendencies in generating PoC formats for various vulnerability types. GPT-4o tends to synthesize executable code-style PoCs (86.3\%), while DeepSeek-R1 focuses more on interactive and manually testable formats (63.4\%).  
}}
\end{tcolorbox}

\begin{figure}[]
  \centering
  \includegraphics[width=0.9\linewidth]{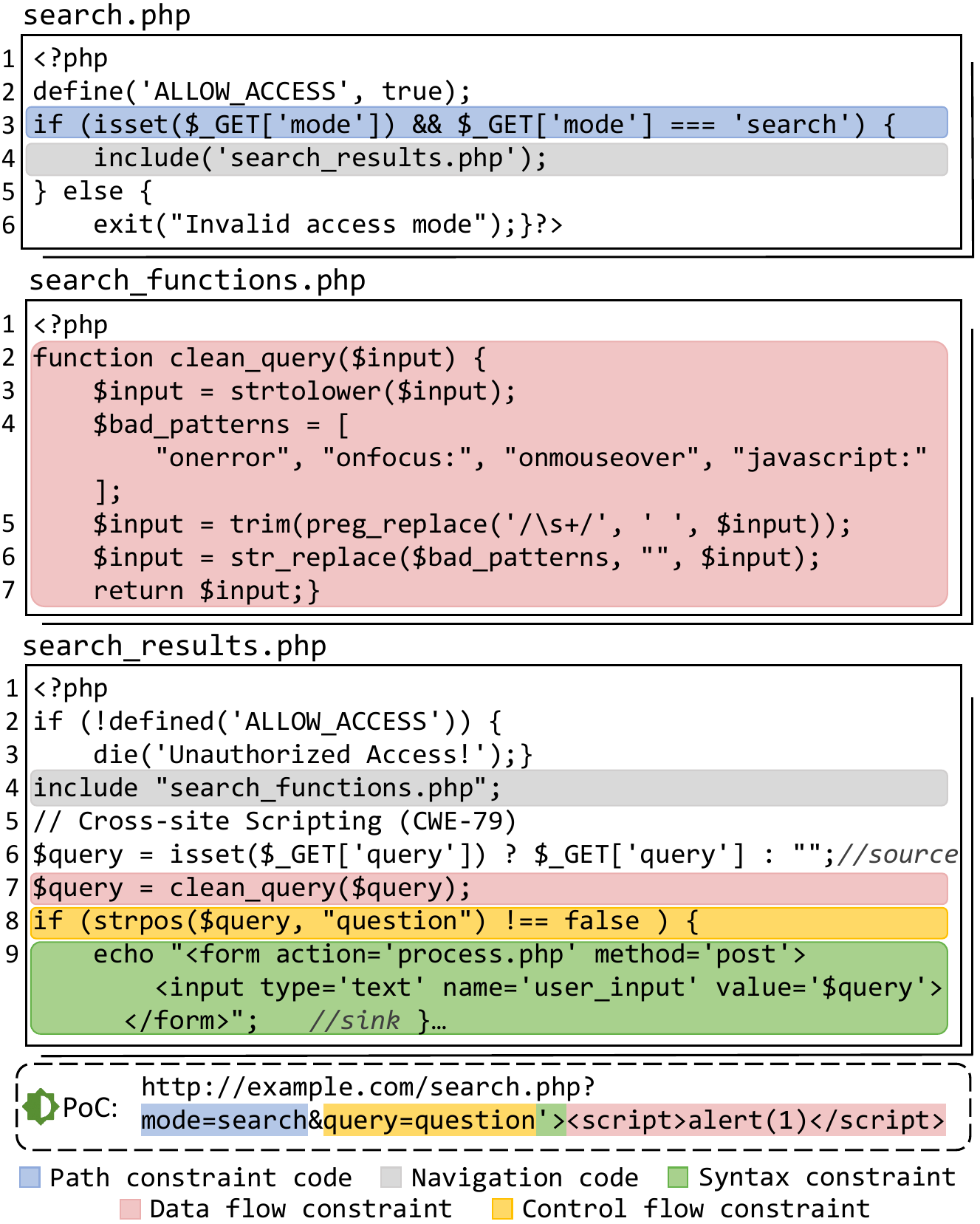}
  \caption{A simplified example of XSS vulnerabilities.}
  \label{fig:example_taint_style}
\end{figure}

\subsection{RQ2: Dissecting the PoC Generation Failures}\label{sec:step2}
We found that 79\% of vulnerabilities still failed to yield functional PoCs. Owing to GPT-4o’s black-box nature, the failure stage is unclear. In this section, we decompose the PoC generation process into distinct sub-tasks by vulnerability type and prompt the LLM to explicitly output each sub-task, allowing precise fault localization without interfering its native reasoning mechanism.

\subsubsection{Setup} 
We reviewed existing PoC generation works and integrated expert knowledge to decompose the PoC generation process for five vulnerability types into three core phases: the \textit{attack vector crafting phase}, the \textit{navigation path generation phase}, and the \textit{PoC assembly phase}.
The navigation path generation phase and PoC assembly phase are consistent across all five vulnerability types, while the key differences lie in the attack vector crafting phase. 
Here, we discuss the attack vector crafting phase for taint-style vulnerabilities, while the corresponding phases for CWE-352 and CWE-434 are presented in detail in Appendix~\ref{Sub-task}.

As shown in~\Cref{fig:example_taint_style}, it briefly shows the PoC generation process for CWE-79 vulnerabilities, which will serve as a running example for detailed analysis throughout the paper. For taint-style vulnerabilities (CWE-78/79/89), we decompose the PoC generation process into 14 sub-tasks, with 7 assigned to the attack vector crafting phase, 4 to the navigation path generation phase, and 3 to the PoC assembly phase.

\noindent\textbf{Attack Vector Crafting Phase.} It consists of 7 sub-tasks and requires the LLMs to model how the \textit{vulnerable variable} propagates from the \textit{source} to the \textit{sink} through \textit{data flow constraints}, \textit{control flow constraints}, and \textit{syntax constraints}, and then reason to generate a valid \textit{attack payload}. As shown in~\Cref{fig:example_taint_style}, LLMs need to first identify the sink (e.g., \texttt{search\_results.php}, Line 9), locate the vulnerable variable, and determine the attacker-controlled source (e.g., \texttt{search\_results.php}, Line 6) to establish the basic elements of the vulnerability. Next, LLMs need to trace the data flow and control flow constraints along the path from the source to the sink. Data flow constraints refer to the data transformation and sanitizer (e.g., \texttt{search\_functions.php}, Lines 2-7), and control flow constraints are the branch conditions imposed by control statements of the vulnerable variable (e.g., \texttt{search\_results.php}, Line 8). In addition, LLMs need to recognize the syntax constraints, which define the contextual syntactic rules that must be complied with at the sink. For example, when the sink is in the HTML tags, one situation is to close the current tag and then embed the attack payload. Another situation is to close the attributes or events of the current tag and embed attack scripts in the new attributes or events, for example, \texttt{" onmouseover=alert(1)}~\cite{buyukkayhan2020s}. Finally, LLMs need to solve an effective payload (e.g., \texttt{question'><script>alert(1)</script>}) that satisfies all data, control, and syntax constraints.

\noindent\textbf{Navigation Path Generation Phase.} It requires the LLM to reason about the execution path from the application's publicly accessible page to the vulnerable sink. It consists of 4 sub-tasks: identifying the \textit{file navigation chain}, \textit{file navigation code}, \textit{path constraint code}, and reasoning \textit{path constraint variables and values}. Simply identifying the sink and generating an attack payload does not guarantee that the vulnerability can be directly reproduced, as the vulnerable module (e.g., the sink) may be deeply embedded within the application and not directly accessible~\cite{alhuzali2016chainsaw, alhuzali2018navex}. As shown in~\Cref{fig:example_taint_style}, \texttt{search\_results.php} enforces file access restrictions on line 2, preventing direct access to the vulnerable phase. Therefore, LLMs need to identify the complete execution path from a publicly accessible page to the sink. LLMs need to first identify the file navigation chain and the corresponding file navigation code. The file navigation chain describes the sequence of file invocations in the application's execution flow, starting from the publicly accessible entry file to the vulnerable file, and file navigation code refers to the specific statements enabling navigation between files (e.g., \texttt{search.php}, Line 4). Next, LLMs need to identify the path constraint code that must be satisfied to reach the file navigation code in each file, meaning local execution paths (e.g., \texttt{search.php}, Line 3). Finally, it needs to solve for the path constraint variables and values, which represent the necessary path inputs conditions (e.g., \texttt{mode=search}) to reach the sink.

\noindent\textbf{PoC Assembly Phase.} It focuses on determining the essential elements required to construct a PoC, including 3 sub-tasks: \textit{request parameters}, \textit{request method}, and \textit{request URL}. 

DeepSeek-R1 outputs the reasoning process when generating PoCs, and we directly analyze its reasoning output to summarize the reasons for failure. For GPT-4o, we design a new prompt that explicitly guides it to output each sub-task result based on public information $\mathcal{P}$ (i.e., $\langle D, P, F \rangle$). As shown in~\Cref{fig:RQ2_Prompt.pdf} in Appendix~\ref{prompts}, this prompt is a simplified version that includes detailed explanations for each sub-task and enforces a standardized JSON output format. Our goal is not to intervene in the model’s internal reasoning, but to pinpoint where and why failures occur. We do not provide any few-shot examples or additional knowledge to avoid introducing unintended biases or external influences on LLM’s responses. 
By analyzing these intermediate outputs, we can further summarize the underlying causes of PoC generation failures identified in \Cref{sec:step1}.

\begin{figure}[]
  \centering
  \includegraphics[width=0.95\linewidth]{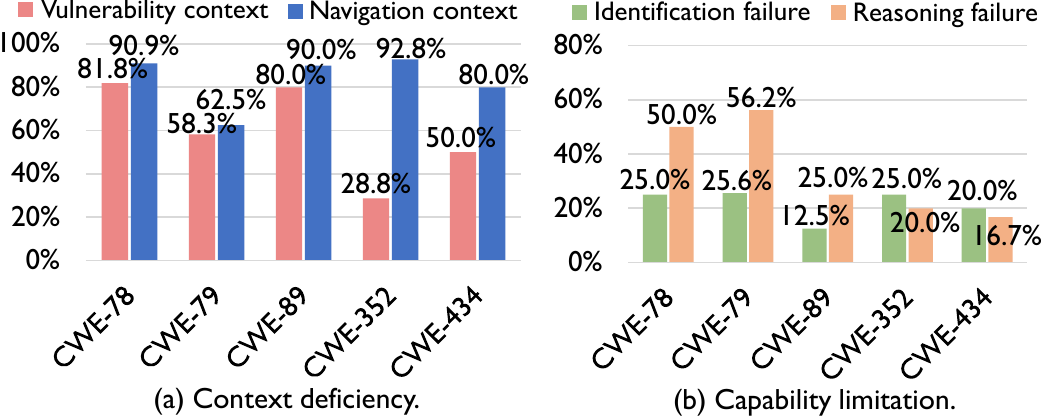}
  \caption{Analysis of failure causes by vulnerability type.}
  \label{fig:RQ2_gpt}
\end{figure}

\subsubsection{Results}
As shown in~\Cref{tab:rq2_all}, we ultimately classified the causes of failures into two categories: Context Deficiency and Capability Limitation. 

\noindent\textbf{C1: Context Deficiency.} 
Complete context comprises two essential components: \textit{vulnerability context} and \textit{navigation context}. The vulnerability context refers to the full contextual details necessary to understand and model the vulnerability. For taint-style vulnerabilities, this includes the source, sink, vulnerable variable, data flow constraints, control flow constraints, and syntax constraints. The navigation context refers to the execution paths and access constraints that lead from publicly accessible pages to the vulnerable sink~\cite{alhuzali2016chainsaw, alhuzali2018navex}. 

As shown in~\Cref{tab:rq2_all}, context deficiency emerges as the dominant cause of PoC generation failures. For GPT-4o, 60.8\% and 81.0\% failures were due to missing vulnerability and navigation context, respectively. For DeepSeek-R1, the corresponding rates were 66.7\% and 86.4\%. We further analyzed the distribution of context deficiency across different vulnerability types and observed distinct tendencies. As shown in~\Cref{fig:RQ2_gpt}a (based on GPT-4o’s failed cases), taint-style vulnerabilities, especially CWE-78 and CWE-89, are more likely to lack key vulnerability context, with rates exceeding 80\%. In CWE-352, navigation context is the most frequently missing, reaching up to 92.8\%. 
For CWE-79, nearly 38\% of the public information $\mathcal{P}$ (i.e., $\langle D, P, F \rangle$) disclosed complete vulnerability and navigation context, and LLM still failed to generate valid PoCs. It suggests that even under ideal disclosure with complete information, inherent capability limitations constrain LLM in complex PoC generation tasks.

\begin{tcolorbox}[size=title,opacityfill=0.1,breakable]
\noindent \textit{\textbf{Finding 5:}} {\textit{Context deficiency is the predominant reason for PoC generation failures. Taint-style vulnerabilities are more prone to missing vulnerability context, whereas CWE-352 are more frequently to suffer from missing navigation context.}}
\end{tcolorbox}

\begin{table}[]
\caption{Analysis of PoC generation failures.}
\centering
\resizebox{1\columnwidth}{!}{
\label{tab:rq2_all}
\begin{tabular}{c|cc|cc}
\hline
\multirow{0}{*}{\textbf{Model}} & \multicolumn{2}{c|}{\textbf{Context Deficiency}}                                                                                                                                & \multicolumn{2}{c}{\textbf{Capability Limitation}}                                                                                                                          \\ \cline{2-5} 
                                & \multicolumn{1}{c|}{\textbf{\begin{tabular}[c]{@{}c@{}}Vulnerability\\ Context\end{tabular}}} & \textbf{\begin{tabular}[c]{@{}c@{}}Navigation\\ Context\end{tabular}} & \multicolumn{1}{c|}{\textbf{\begin{tabular}[c]{@{}c@{}}Ident. Failure\end{tabular}}} & \textbf{\begin{tabular}[c]{@{}c@{}}Reason. failure\end{tabular}} \\ \hline
\textbf{GPT-4o}                          & \multicolumn{1}{c|}{60.8\%}                                                                       & 81.0\%                                                                    & \multicolumn{1}{c|}{24.3\%}                                                                  & 34.9\%                                                                \\
\textbf{DeepSeek-R1 }                    & \multicolumn{1}{c|}{66.7\%}                                                                       & 86.4\%                                                                    & \multicolumn{1}{c|}{18.9\%}                                                                  & 33.8\%                                                                \\ \hline
\end{tabular}
}
\end{table}

\noindent\textbf{C2: Capability Limitation.} 
We further categorize the LLM capabilities into two dimensions: \textit{identification capability and reasoning capability.} 
To quantify how these capabilities lead to failure, we evaluate the effectiveness of LLMs on identification sub-tasks and reasoning sub-tasks during the generation process. 
The quantification metrics are defined as:  
\begin{equation}
\text{Ident. failure} = 1 - \frac{1}{2} \left( 
\frac{N_{\text{Ident\_vul}}}{N_{\text{Complete\_vul}}} 
+ 
\frac{N_{\text{Ident\_nav}}}{N_{\text{Complete\_nav}}} 
\right)
\label{eq:ide}
\end{equation}

\begin{equation}
\text{Reason. failure} = 1 - \frac{1}{2} \left( 
\frac{N_{\text{Gen\_payload}}}{N_{\text{Ident\_vul}}} 
+ 
\frac{N_{\text{Gen\_pathvar}}}{N_{\text{Ident\_nav}}} 
\right)
\label{eq:rea}
\end{equation}
where \( N_{\text{Complete\_vul}} \)/\( N_{\text{Complete\_nav}} \) indicates total vulnerabilities with complete vulnerability/navigation contexts, \( N_{\text{Identify\_vul}} \)/\( N_{\text{Identify\_nav}} \) indicates total vulnerabilities where all required vulnerability/navigation sub-tasks are correctly identified, and  
\( N_{\text{Gen\_payload}} \)/\( N_{\text{Gen\_pathvar}} \) indicates total vulnerabilities that successfully generated payload/path constraint variables and values. 

As shown in~\Cref{tab:rq2_all}, the identification and reasoning capabilities of LLMs differ, with DeepSeek-R1 achieving better overall effectiveness than GPT-4o.
As shown in~\Cref{fig:RQ2_gpt}b, we further assess GPT-4o’s capability limitation across vulnerability types and find it demonstrating stable identification capabilities with all five types achieving above 74\%, whereas its reasoning ability varies significantly. 
Specifically, GPT-4o shows strong reasoning ability on CWE-434 and CWE-352 (83.3\% and 80.0\%), and reasonably on CWE-89 (75.0\%), but shows notable degradation on CWE-78 and CWE-79, with failure rates reaching 50.0\% and 56.2\%, respectively. 
This may be attributed to the complex constraints required in attack payload construction, exceeding the model’s shallow reasoning capacity under basic prompt (see~\Cref{fig:RQ2_Example.pdf}). In other words, when facing multi-step, syntax- and semantics-sensitive PoC synthesis tasks, LLMs require more guided prompting strategies to unlock their capabilities.
 
As DeepSeek-R1 has relatively fewer vulnerabilities with complete context (e.g., only one case in CWE-89), we do not further quantify its capabilities across vulnerability types to avoid skewed analysis.

\begin{tcolorbox}[size=title,opacityfill=0.1,breakable]
\noindent \textit{\textbf{Finding 6:}} {\textit{With complete contextual information, the LLM demonstrates stable identification capability for different vulnerability types, consistently exceeding 74\%. However, its reasoning capability remains suboptimal when handling CWE-78 and CWE-79, not exceeding 50\%.
}}
\end{tcolorbox}

\subsection{RQ3: Enhancement Through Context Supplement}\label{sec:step3}
Building on RQ2's findings that insufficient contextual information critically hinders LLM's capabilities on PoC generation, this section investigates how supplementing missing vulnerability context (e.g., data flow constraints) and navigation context (e.g., file hierarchy dependencies) impacts PoC synthesis effectiveness. 
To isolate the influence of context granularity, we design a controlled experiment where we manually reconstruct these contexts at two distinct levels (file-level and function-level, separately).

\subsubsection{Setup} We {manually} extract vulnerability and navigation context from source code {of our benchmark} through program analysis at both file-level and function-level granularity.

\noindent\textbf{Vulnerability Context Extraction.} We first introduce the extraction for \textit{file-level vulnerability context}. We perform initial analysis based on the vulnerability description $D$ and patch commit $P$ to extract all files involved in the propagation path from the attacker-controlled source to the sink. 
Specifically, we trace variables in $D$ and $P$, and then track assignments and data dependencies associated with them. 
If the data dependency path involves function parameter passing, we conduct a detailed analysis of the function call relationships, map parameters to variables, and continue tracking within the calling function~\cite{alhuzali2018navex}. 
If the data dependency path includes global variables, we employ a global search to locate all statements that may use the variable and proceed with the analysis. 
Using these data flow analysis rules, we perform forward taint propagation to identify sinks. 
The identification of sink statements leverages previous work modeling various PHP built-in sinks for different vulnerability types~\cite{backes2017efficient,huang2019uchecker,huang2021ufuzzer}, supplemented by expert knowledge. For CWE-352, we prioritize semantic sinks (e.g., deleteAll()) over syntactic patterns, aligning with exploit logic analysis.
We confirm the variable associated with identified sink statements as the vulnerability variable and perform backward taint analysis to locate the source directly controlled by the attacker~\cite{shi2024recurscan}. 
All files involved in the data flow path are included to construct the file-level vulnerability context. Regarding extracting the \textit{function-level vulnerability context}, we adopt a similar strategy to extract all relevant functions involved in the vulnerability path. 
Additionally, if the data dependency statement is a global statement outside of any function, we consider the entire file as a single function and exclude extraneous functions and control structures. For statements within HTML modules, we treat the innermost enclosing tag and its content as a single function.

\noindent\textbf{Navigation Context Extraction.} We begin with the vulnerability context and perform a backward analysis to identify the first publicly accessible page.
All relevant files and constraints encountered during this process are collected to form the \textit{file-level navigation context}. A publicly accessible page is defined as one where, along the navigation path from this page to the vulnerable sink, all assignments to superglobal variables and navigation-related local variables satisfy their respective constraints. If superglobal variables are undefined, the program will terminate prematurely~\cite{alhuzali2016chainsaw}; if navigation-related local variables do not meet the constraints, the program will not execute according to the intended workflow~\cite{alhuzali2018navex}. Based on this, we start from the vulnerability context, adding it to the navigation context and checking if it is publicly accessible. If it is, we output the navigation context directly. If not, we identify statements in the code responsible for navigation (e.g., redirection, link, include, etc.) to locate files that can reach the vulnerable component, add these files to the navigation context, and continue to check their accessibility. This process is repeated until we identify the first publicly accessible page. All involved files are then included in the final file-level navigation context. When extracting \textit{function-level navigation context}, we adopt a similar strategy to extract all relevant functions involved in the navigation path.

To evaluate the effectiveness of directly synthesizing PoCs under sufficient information, we provide LLMs with the vulnerability description $D$, patch commit $P$, complete vulnerability and navigation context, and the public access entry URL. 
The prompt template is shown in~\Cref{fig:RQ3_Prompt.pdf} in Appendix~\ref{prompts}. 
For vulnerabilities that failed due to context deficiency, we supplement complete vulnerability and navigation context at file-level and function-level, separately. 
For vulnerabilities where the public information $\mathcal{P}$ is complete  (i.e., $\langle D, P, F \rangle$) but the failure is due to LLM capability limitations, we further refine the vulnerable file to the function-level. 
Since LLMs have a maximum context size (i.e., DeepSeek-R1 supports up to 64K tokens and GPT-4o supports up to 128K tokens), file-level context may exceed this limit.
If the total input surpasses the model's size, we remove some functions unrelated to the vulnerability to ensure the remaining context fits within the token limit. 
We conducted three independent trials for each input, using the same evaluation criteria as in~\Cref{sec:step1}.

\begin{figure}[]
  \centering
  \includegraphics[width=0.8\columnwidth]{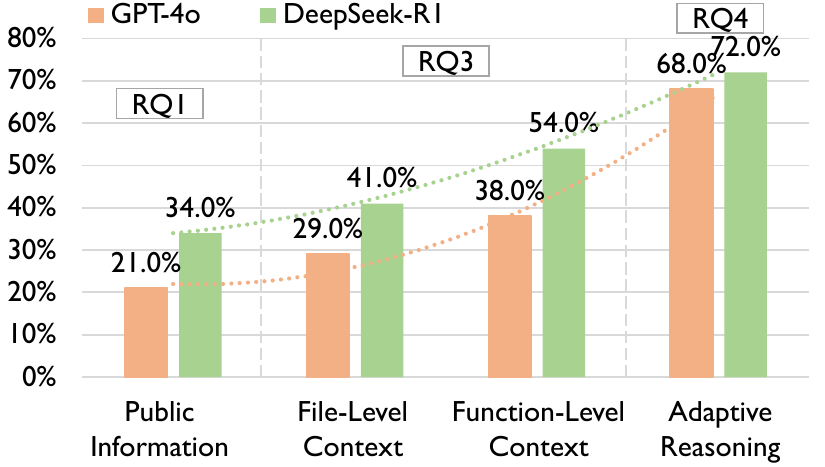}
  \caption{Overall PoC generation success rates.}
  \label{fig:RQ1_RQ3_RQ4_all}
\end{figure}

\subsubsection{Results.} 
As shown in the~\Cref{fig:RQ1_RQ3_RQ4_all}, the PoC generation capability of LLMs improves with the supplementation of both vulnerability and navigation contexts, with function-level yielding greater gains (9-13\%) than file-level. 
This reflects the sensitivity of LLMs to input granularity, especially in code-intensive settings, where excessive surrounding context may distract the LLMs from capturing the vulnerability semantics and exploitation reasoning due to limited attention capacity~\cite{tian2023chatgpt}.  
Furthermore, with complete function-level context, the reasoning LLM (DeepSeek-R1) achieves a 54\% success rate, demonstrating strong capabilities in handling structured, multi-step reasoning tasks. 
In contrast, GPT-4o still shows a limited effectiveness, only 38\% under identical context conditions, indicating continued challenges in synthesizing PoCs for complex vulnerabilities that require multi-step logic or strict constraint satisfaction. 
This effectiveness gap shows GPT-4o's bottleneck shifts from contextual granularity to its inherent reasoning limitations. 
These findings suggest that direct synthesis using basic prompts remains insufficient, and further improvements such as incorporating chain-of-thought (CoT) prompting and in-context learning (ICL) are essential to enhance reasoning depth and capabilities. 
This observation motivates an investigation into whether advanced prompt engineering strategies can effectively enhance effectiveness.

\begin{tcolorbox}[size=title,opacityfill=0.1,breakable]
\noindent \textit{\textbf{Finding 7:}} {
\textit{Supplementing complete vulnerability and navigation context improves PoC generation effectiveness to 54\% for DeepSeek-R1 and 38\% for GPT-4o, with function-level providing 9\%-13\% improvement than file-level ones.}}
\end{tcolorbox}

\begin{figure}[t]
  \centering
  \includegraphics[width=1\linewidth]{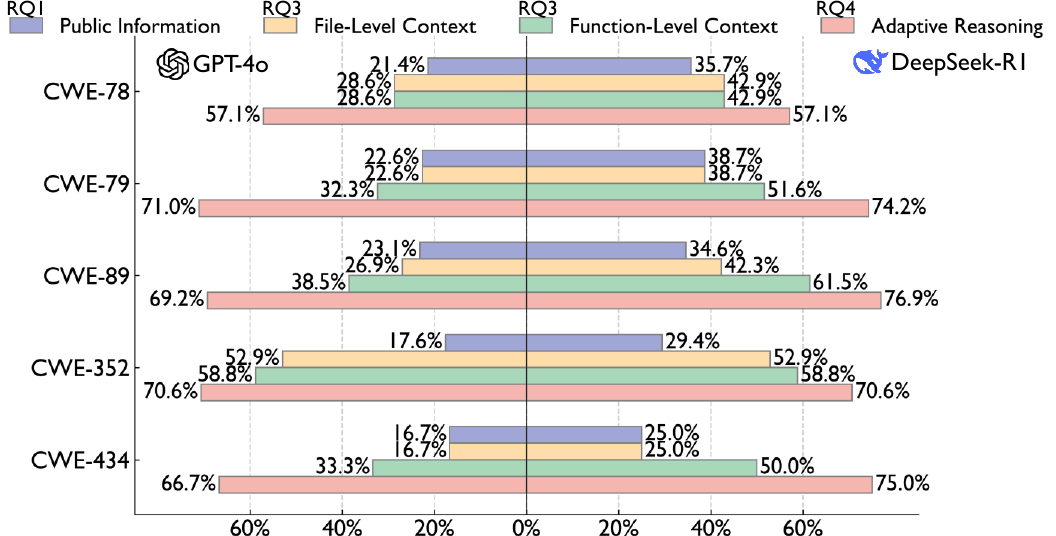}
  \caption{PoC generation success rates of different vulnerability types.}
  \label{fig:RQ1_RQ3_RQ4_detai}
\end{figure}

As shown in the~\Cref{fig:RQ1_RQ3_RQ4_detai}, when supplementing function-level context, GPT-4o and DeepSeek-R1 demonstrate improved PoC generation effectiveness across all of these vulnerability types, with the most notable gains observed in CWE-352 (41.2/29.4\%), CWE-89 (15.4/26.9\%), and CWE-434 (16.6/25.0\%). In contrast, the improvements for CWE-78 (7.2/7.2\%) and CWE-79 (9.7/12.9\%) were relatively limited. 
Although CWE-78, CWE-79, and CWE-89 all fall under taint-style vulnerabilities, they exhibit different responses to context supplementation.  
For CWE-89, even if a payload does not result in direct data exfiltration or manipulation, triggering an SQL syntax error can still be considered a successful PoC~\cite{guler2024atropos}, which reduces the difficulty of payload generation. 
For CWE-78, the vulnerable variables often undergo multiple transformations before reaching the sink, increasing the burden on LLMs to model data flow constraints and navigation path. 
CWE-79 is constrained by strict syntax requirements at the sink, such as escaping or embedding payloads into specific HTML contexts, making it more difficult to construct valid and executable attack payloads.
\begin{tcolorbox}[size=title,opacityfill=0.1,breakable]
\noindent \textit{\textbf{Finding 8:}} {\textit{
Context supplementation yields substantial gains for GPT-4o/DeepSeek-R1 on CWE-352 (41.2/29.4\%), CWE-434 (16.6/25.0\%), and CWE-89 (15.4/26.9\%), but provides limited improvements on CWE-78 (7.2/7.2\%) and CWE-79 (9.7/12.9\%).
}} 
\end{tcolorbox}

\subsection{RQ4: Enhancement Through Adaptive Reasoning}\label{sec:step4}
Our aforementioned experiments reveal that while LLMs exhibit foundational capabilities for generating PoCs, their effectiveness degrades when confronting vulnerabilities that require multi-stage exploit chaining or precise constraint satisfaction. Moreover, PoC generation is a long-chain reasoning task, where errors in intermediate steps can propagate and ultimately invalidate the output.
Therefore, we explore prompt engineering's impact on PoC generation. We employ five adaptive reasoning prompts, including CoT and ICL techniques, to guide the LLM to perform step-by-step reasoning and a real-time feedback mechanism to refine outputs iteratively. 

\subsubsection{Setup}
\begin{figure}[t]
  \centering
  \includegraphics[width=0.9\linewidth]{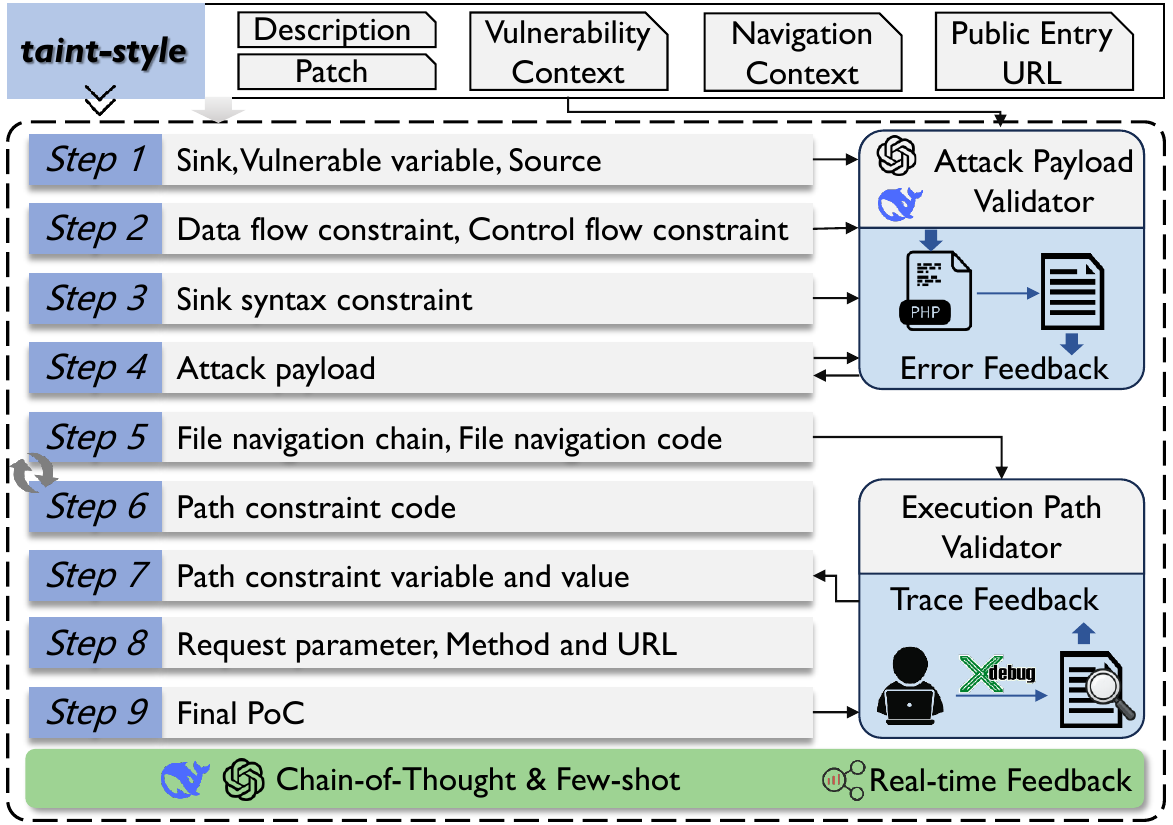}
  \caption{Overview of taint-style vulnerability PoC generation.}
  \label{fig:RQ4_taintstyle.pdf}
\end{figure}

As illustrated in~\Cref{fig:RQ4_taintstyle.pdf}, we present our adaptive reasoning prompt design tailored for taint-style vulnerabilities, including CoT reasoning, in-context learning, and real-time feedback validators. Similar tailored strategies (e.g., CoT, ICL, and validators) are applied to CWE-352 and CWE-434, though we omit the detailed design here for brevity.

\noindent\textbf{Chain-of-Thought.} LLMs’ reasoning capabilities can be significantly enhanced using the $\langle$input, \textit{CoT}, output$\rangle$ prompting format~\cite{wei2022chain}. 
A \textit{CoT} here is a series of intermediate natural language reasoning steps that lead to the outcome. 
To implement step-by-step reasoning, we decompose the PoC generation process based on the sub-tasks defined in~\Cref{sec:step2}. As shown in~\Cref{fig:RQ4_taintstyle.pdf}, we originally split the taint-style vulnerabilities PoC generation process into 14 sub-tasks. However, solving each sub-task individually could lead to redundant reasoning and increase the token burden for LLMs. To optimize efficiency, we merge interdependent tasks, reducing the process to 9 key steps while ensuring PoC completeness and coherence. 

\noindent\textbf{In-context Learning.} Our previous experiments revealed that certain sub-tasks require specialized PoC knowledge, such as constructing attack payloads in ~\Cref{sec:step2}. 
To address this, we apply the in-context learning~\cite{wei2022chain} technique by providing few-shot examples tailored to certain sub-tasks to enhance the reasoning ability. 
We collect sub-task representations from related research, GitHub repositories, and real-world vulnerability reports. 
For all vulnerability types, we provide few-shot examples for \textit{file navigation code} to help LLMs understand the relationship between different files. Specifically, for taint-style vulnerabilities, we provide few-shot examples for \textit{sink}, \textit{source}, \textit{data flow constraints}, \textit{syntax constraints} and \textit{attack payload}~\cite{alhuzali2018navex,stasinopoulos2019commix,shi2024recurscan, buyukkayhan2020s,sqlpayloadb, payloadb93:online}. 
For CWE-352 vulnerabilities, we include examples for CSRF \textit{protection mechanisms} and \textit{bypass techniques}~\cite{sudhodanan2017large, likaj2021we}. Similarly, for CWE-434 vulnerabilities, we provide examples covering file \textit{validation mechanisms}, \textit{bypass techniques} and \textit{upload code}~\cite{lee2020fuse, huang2019uchecker, huang2021ufuzzer}. Other sub-tasks are highly dependent on specific vulnerability instances, and we do not provide examples.

We design a real-time feedback mechanism that uses validators to verify responses and collect real-time feedback to iteratively refine PoC generation. 
As shown in~\Cref{fig:RQ4_taintstyle.pdf}, we design two validators: the attack payload validator and the execution path validator, which address the core challenges of PoC generation for taint-style vulnerabilities. 

\begin{figure}[t]
  \centering
  \includegraphics[width= \linewidth]{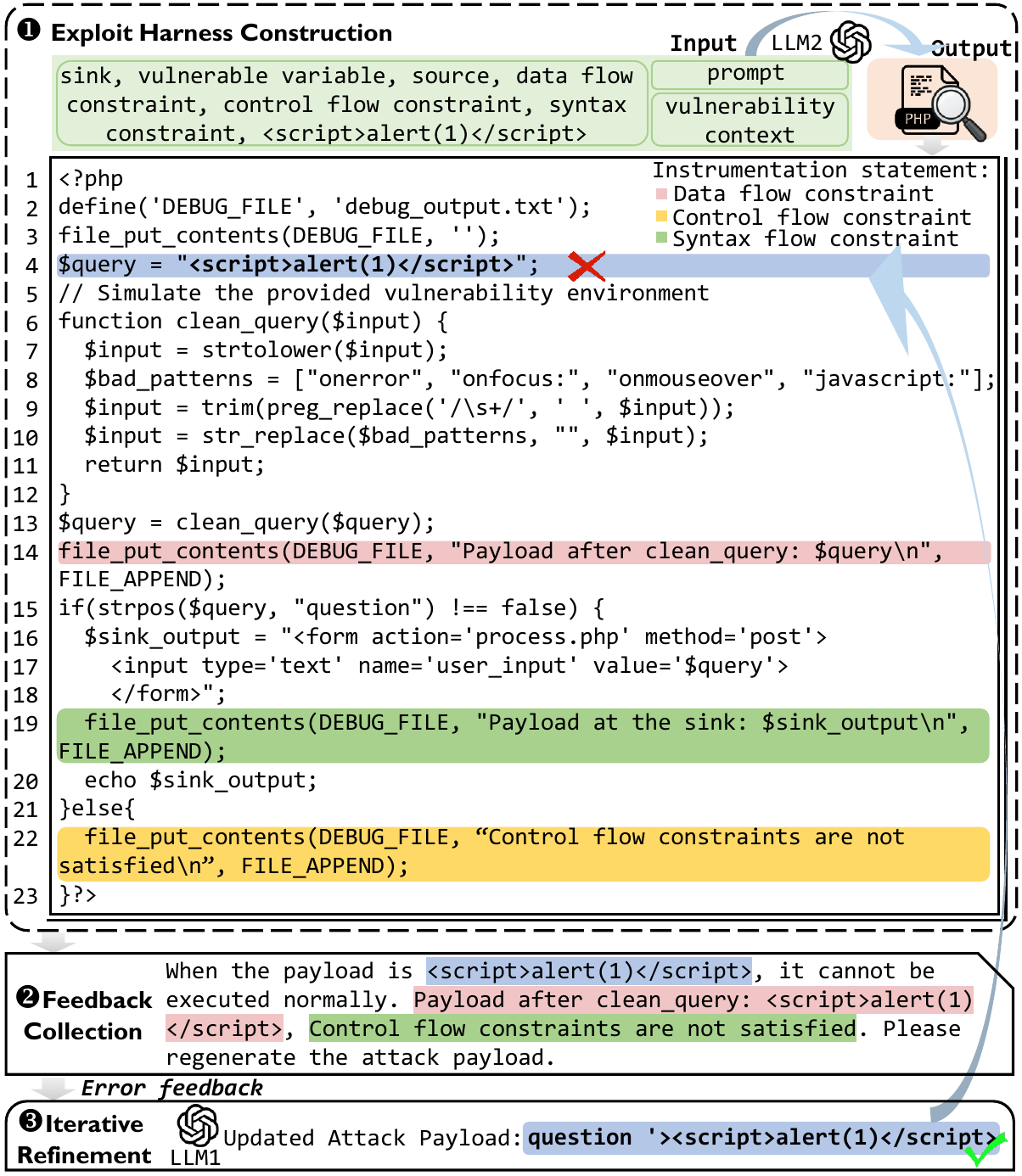}
  \caption{The example of adjusting the attack payload.}
  \label{fig:img/RQ4_payload}
\end{figure}

\noindent\textbf{Attack payload validator.} As shown in~\Cref{fig:img/RQ4_payload}, the attack payload validator is used to verify the vulnerability example in~\Cref{fig:example_taint_style} of~\Cref{sec:step2}. It operates in three stages: (1) exploit harness construction, (2) feedback collection, and (3) iterative refinement. The validator checks whether the original payload (e.g., \texttt{<script>alert(1)</script>}) generated by the LLM is valid for the target vulnerability. If not, constraint-based feedback is collected and returned to the initial LLM to iteratively refine the output (e.g., \texttt{question '><script>alert(1)</script>}).

In the exploit harness construction stage, a new LLM instance is invoked to synthesize a payload verification environment that captures three runtime constraint details associated with the payload execution. To enable this, we first gather the \textit{sink}, \textit{vulnerability variable}, \textit{source}, \textit{data flow constraints}, \textit{control flow constraints}, \textit{syntax constraints}, the currently generated \textit{attack payload} from the initial LLM (denoted as \texttt{LLM1}), and complete the function-level \textit{vulnerability context}. Then, we invoke a separate LLM instance (denoted as \texttt{LLM2}) and design prompts to guide it in constructing a localized PHP vulnerability verification environment based on this information to validate the effectiveness of the payload, as shown in~\Cref{fig:img/RQ4_payload} (Line 1-23). 
Since the payload must satisfy data flow, control flow, and syntax constraints, we also instruct the \texttt{LLM2} to insert instrumentation statements into the environment to capture constraint-specific execution details: 

\begin{itemize}[leftmargin=5pt]
\item Control flow constraint. Output ``control flow constraints are not satisfied.'' if the payload fails to reach the sink, as shown in~\Cref{fig:img/RQ4_payload} (Line 22). This helps determine whether the payload satisfies the necessary control branch conditions to reach the final sink.
\item Data flow constraint. Output value of payload after it passes through the data transformation and sanitization function, as shown in~\Cref{fig:img/RQ4_payload} (Line 14). This helps assess whether the payload remains effective after data flow sanitization.
\item Syntax constraint. Output actual representation of payload at the sink, as shown in~\Cref{fig:img/RQ4_payload} (Line 19). This helps verify whether the payload is syntactically valid and executable.
\end{itemize}

In the feedback collection stage, we execute the localized vulnerability environment and determine whether the attack payload is generated successfully. If not, we provide the following error feedback to the \texttt{LLM1}: (1) Payload usability. We explicitly tell the LLM that the current payload is not available. (2) Error feedback. Since our execution environment automatically collects real-time feedback on three constraints, we provide this information to LLM.

In the iterative refinement stage, guided by the error feedback, the \texttt{LLM1} refines its output and regenerates a new payload (\texttt{question '><script>alert(1)</script>}), which is subsequently validated again within the PHP environment. To prevent \texttt{LLM1} from entering an infinite adjustment loop, we limit the maximum feedback iterations for each validator to 3 to balance thoroughness and efficiency~\cite{zhang2024acfix}.

\noindent\textbf{Execution Path Validator.} LLMs may generate an incorrect navigation path, leading to PoC failure. Therefore, we introduce the execution path validator, as shown in~\Cref{fig:RQ4_taintstyle.pdf}, to provide actual execution feedback through dynamic trace information recorded by Xdebug to refine the path generation. 
Xdebug is an open-source PHP debugger that monitors requests during program execution and records the corresponding call stack information to local trace files~\cite{XdebugDe65:online}. 
Specifically, we run the generated PoC in a local reproduction environment and obtain an execution trace file. If the PoC fails, we parse the trace file and extract the file and function call information, comparing them with the \textit{file navigation chain} and \textit{file navigation code} identified by LLM to detect untriggered file navigation chain nodes. We then provide LLM with the following trace feedback to update path constraints and values and reconstruct PoCs: (1) PoC usability. We explicitly inform the LLM that the PoC failed. (2) Trace feedback. We provide feedback information: ``The current PoC fails to fully execute the file navigation chain, and application execution flow failed to reach \textless \textit{untriggered file node}\textgreater\ or \textless \textit{sink}\textgreater''. 

We provide LLMs with the vulnerability description, patch commit, complete function-level vulnerability, and navigation context, as well as the public entry URL, and apply the above adaptive reasoning prompt strategies to explore their potential.

\subsubsection{Results} 
As shown in~\Cref{fig:RQ1_RQ3_RQ4_all}, adaptive reasoning prompt strategies significantly improve PoC generation effectiveness, with DeepSeek-R1 achieving 72.0\%, slightly outperforming GPT-4o at 68.0\%. Despite GPT-4o's initial performance lag, it benefits more from prompt techniques—with a 30.0\% improvement—compared to DeepSeek-R1’s 18.0\% gain.
This improvement indicates that GPT-4o’s effectiveness is more sensitive to external guidance. 
In particular, it also validates the effectiveness of our curated CoT prompts and sub-task decomposition design, which successfully enhances reasoning in generic models. 
In contrast, DeepSeek-R1 demonstrates strong baseline performance even without prompt optimization, achieving 54.0\% with supplementing function-level context (\Cref{sec:step3}).
Its ability to autonomously decompose the PoC generation process into reasoning steps, perform validation, and engage in self-reflection reflects its built-in reasoning strength. 
While CoT prompts offer to further regularize its reasoning process, their marginal impact suggests that DeepSeek-R1’s effectiveness is primarily limited by input completeness rather than reasoning limitations.

\begin{tcolorbox}[size=title,opacityfill=0.1,breakable]
\noindent \textit{\textbf{Finding 9:}} {\textit{
Adaptive reasoning prompt strategies effectively unlock the PoC generation potential of both models, with DeepSeek-R1 achieving a final success rate of 72\% and GPT-4o reaching 68\%. 
}}
\end{tcolorbox}

As shown in~\Cref{fig:RQ1_RQ3_RQ4_detai}, adaptive reasoning prompt strategies lead to obvious improvements across CWE-78 (28.5/14.2\%), CWE-79 (38.7/22.6\%), CWE-89 (30.7/15.4\%), and CWE-434 (33.4/25.0\%) for GPT-4o and DeepSeek-R1, respectively. The gain for CWE-352 is relatively small, with both models only improving by 11.8\%. 
This modest improvement can be attributed to its nature, which typically require constructing a forged HTTP request. 
Once the LLM is provided with sufficient information, the core challenge shifts to accurately capturing the request parameters and execution path. 
However, when the application uses complex routing mechanisms to control execution flow, even with step-by-step path reasoning and using an execution path validator, the improvement remains limited. LLMs still struggle to infer the real execution path in scenarios involving complex dynamic module transitions and parameter bindings. 
In addition, we further observe that both models perform significantly worse on CWE-78, achieving only a 57.1\% success rate, while they reach or approach 70\% on other types. 
This is primarily due to the complex exploit chains involved in CWE-78, where the payload must pass through multiple function calls, parameter transformations, string operations, sanitization, and condition checks before reaching the final sink. 
These deep and layered path dependencies introduce strict data and control flow constraints that LLMs often fail to fully capture.

\begin{tcolorbox}[size=title,opacityfill=0.1,breakable]
\noindent \textit{\textbf{Finding 10:}} {\textit{Adaptive reasoning prompts can effectively improve LLMs' PoC generation capabilities for most vulnerability types. 
However, their reasoning effectiveness remains limited when handling long exploit chains or complex dynamic routing logic.
}}
\end{tcolorbox}

We further summarize the overall time and monetary expenditures of both LLMs in our empirical study. While DeepSeek-R1 required 3.35 times longer than GPT-4o (47.06 hours vs. 14.02 hours), it reduced the cost by a factor of 18.65 (2.02 USD vs. 37.68 USD).

\section{Discussion}
\subsection{Lessons Learned}
We summarize the current state quo of LLM-based PoC generation and identify several open challenges and future research directions: 

\subsubsection{Vulnerability Type Extension}
We conducted an initial exploration of LLM-driven PoC generation in web applications and achieved a 68\%-72\% success rate in~\Cref{sec:step4}, showing the potential of LLMs. 
While our study primarily focuses on five representative CWE types, the underlying task decomposition and prompt design are extensible to other vulnerability classes by incorporating their specific exploitation logic.
However, as exemplified by CSRF and taint-based vulnerabilities in~\Cref{sec:step2}, different vulnerability types may exhibit substantial divergence in their exploitation mechanisms. Such divergence necessitates a pre-understanding of the vulnerability landscape in order to appropriately deconstruct the task into meaningful sub-tasks. Furthermore, maintaining the effectiveness of these sub-tasks often entails considerable effort and iterative refinement. 
Despite the solid foundation and practical guidance established by our methodology and empirical results, adapting LLMs to the exploitation patterns of other vulnerability types remains a challenge that warrants further in-depth investigation.

\subsubsection{Knowledge Augmentation} The completeness of input information has a direct impact on the effectiveness of PoC generation. Increasing the richness of the input leads to better results. For instance, in~\Cref{sec:step1}, using only the vulnerability description resulted in a PoC generation success rate of 8\%. When we incorporated additional elements such as patch commits, vulnerability context, and navigation context, the success rate improved significantly to 38\%–54\%, even without applying any prompt optimization strategies. In practice, disclosed vulnerabilities are often accompanied by abundant auxiliary information—such as discussions, comments, and threat intelligence—which can be extracted and structured to further enhance PoC generation. However, as shown in~\Cref{sec:step3}, introducing excessive or irrelevant context may lead to increased noise, dilute the model’s attention, and ultimately degrade performance. Therefore, developing heuristics strategies for the selective collection and extraction of relevant information is a crucial direction for future research.

\subsubsection{Reasoning Bottleneck} We observed that the reasoning accuracy of LLMs significantly degrades when the vulnerability path involves multiple function calls, complex conditional branches, or data transformations, as shown in~\Cref{sec:step4}. This is primarily due to the probabilistic nature of LLM generation, which lacks precise execution capabilities. In such cases, LLMs often struggle to accurately track variable states, maintain logical consistency across execution paths, or infer implicit constraints introduced by program semantics. These limitations indicate that, in complex vulnerability scenarios, it may be necessary to further integrate dynamic analysis or static analysis like symbolic execution techniques to augment the reasoning capabilities of LLMs.

\subsubsection{End-to-End Automation}
In~\Cref{sec:step4}, the core PoC generation is automated via LLMs, while real-time feedback mechanism remains minimal manual due to practical challenges.
For example, in the attack payload validator, although the LLM can synthesize a validation environment to gather constraint information, it is still up to human experts to determine—based on domain knowledge—whether the payload was successful. Similarly, in the execution path validator, manual execution of the PoC is required to collect trace information for further refinement. 
Interactive feedback remains difficult to automate due to high false positive rates in behavioral heuristics and the diversity of PoC formats and runtime environments.
The automated reproduction of PoCs and the reliable validation of their outcomes remain open challenges.

\subsection{Threats to Validity}

The first threat concerns potential LLM data leakage. To assess this, we prompted only a CVE-ID and observed that the generated PoCs mismatched the actual vulnerabilities in the affected software, type, and exploitation method, indicating negligible influence from memory retrieval. 
Additionally, some raw PoCs in our dataset exist in unstructured forms (e.g., images or videos), which are unlikely to be effectively indexed during pretraining.

The second potential threat comes from the generalizability of data and model selection. 
Reproducing real-world vulnerabilities is a resource-intensive task, which required 600+ man-hours to build 62 isolated environments and 121+ man-hours for validating 2,900+ PoCs (across RQ1/3/4) of 100 cases.
To mitigate it, we select five high-impact web vulnerability types in PHP applications and construct a dataset consisting of recent, high-severity, and representative vulnerabilities, which captures diverse and critical vulnerability patterns. Regarding LLM selection, we selected two kinds of SOTA LLMs: GPT-4o, a representative general-purpose model, and DeepSeek-R1, a representative reasoning-based LLM.

The third potential threat arises from the empirical sub-task decompositions and the designed prompts. To mitigate it, two authors independently decompose the overall task into sub-tasks based on relevant literature and their experience. They reached an agreement through collaborative discussions. After finalizing the sub-task definitions, we developed a stepwise prompt strategy tailored to each sub-task. These strategies were iteratively refined on a set of vulnerability cases to achieve optimal performance.

The fourth potential threat arises from bias in the manual reproduction of vulnerabilities and file and function level context construction. To mitigate this, we employed a consensus-based evaluation strategy: two authors independently reviewed all outputs, and any disagreements were resolved through discussion to reach a consensus.

\section{Related Work}

\subsection{PoC Generation for Web Applications}
Automated PoC generation for web vulnerabilities has long relied on traditional program analysis techniques~\cite{kieyzun2009automatic, huang2013craxweb, alhuzali2016chainsaw, pellegrino2017deemon, alhuzali2018navex, zhao2022cefuzz, lee2020fuse, chen2023uradar, zhao2023remote, park2022fugio, bensalim2021talking, steffens2020pmforce, xiao2021abusing, cassel2025nodemedic, kang2023scaling}. Seminal works like NAVEX~\cite{alhuzali2018navex} combined static/dynamic analysis for exploit synthesis, URadar~\cite{chen2023uradar} generates PoCs based on predefined rules, and Cefuzz~\cite{zhao2022cefuzz} and FUGIO~\cite{park2022fugio} employed fuzzing for specific vulnerabilities. These methods require laborious manual pattern engineering and struggle to scale with modern web applications' logic and semantic complexity.
Recent advances in LLMs present potential for this task, yet systematic explorations remain strikingly absent. 
Recent work by Fang et al.~\cite{fang2024llm} conducts an initial exploration of LLMs for PoC generation, demonstrating feasibility on 15 CVEs through basic prompts. While this preliminary effort suggests LLMs' potential in processing CVE descriptions for PoC generation, their work suffers three limitations: 
(1) Scope and Methodology: Vulnerability selection lacks transparency (types/criteria unstated), and single-dimension analysis (text-only inputs) ignores code-level semantics critical for web vulnerabilities; 
(2) Depth-Generality Tradeoff: With limited cases and no task decomposition, observed results may reflect dataset-specific biases rather than LLMs' intrinsic capabilities;
(3) Reproducibility: Undisclosed prompts and validation steps hinder independent verification.
By addressing these limitations through scaled experimentation (100 reproducible CVEs and 3,000 trials), four-stage hierarchical evaluation, and open artifact release, we establish the first foundation for LLM-driven PoC generation—revealing both capabilities previously obscured by methodological constraints and fundamental challenges requiring future research.

\subsection{PoC Generation for Binaries Vulnerabilities}
Some existing works focus on binary programs~\cite{brumley2008automatic,cha2012unleashing,avgerinos2014automatic,xu2018automatic,you2017semfuzz,lee2021constraint,yang20231dfuzz} and generate PoCs for memory vulnerabilities, can be categorized into two directions: symbolic execution~\cite{brumley2008automatic,cha2012unleashing,avgerinos2014automatic,xu2018automatic} and directed fuzzing~\cite{you2017semfuzz,lee2021constraint,yang20231dfuzz}. 
For example, APEG~\cite{brumley2008automatic} collected and solved constraints on paths reachable to the patch to construct PoCs for input validation vulnerabilities. 
On the other hand, Semfuzz~\cite{you2017semfuzz} extracted semantic knowledge from the vulnerability-related text and guided the fuzzing to generate PoCs of Linux kernel bugs.
However, these work primarily focuses on traditional memory vulnerability types and techniques. 
In contrast, our study focuses on web applications, which introduces new challenges beyond memory vulnerabilities. Our study offers a novel and effective direction for future research on LLM-driven PoC generation. 

\section{Conclusion}
In this paper, we conducted a systematic empirical analysis for LLM-driven PoC generation for web vulnerabilities. Through an evaluation of 100 reproducible CVEs, we demonstrate that while LLMs can generate functional PoCs from public information (8-34\% success), their effectiveness is fundamentally constrained by missing contextual information. 
Our phased analysis reveals actionable pathways for improvement: 
supplementing function-level context, adaptive prompting engineering using CoT and ICL reasoning and real-time feedback, effectively bridging the capability gap between general and reasoning LLMs. 
In addition, 14 PoCs generated by LLMs have been accepted by NVD and 9 PoCs by Exploit DB, further reinforcing the practical relevance of our work. 
LLMs can synthesize PoCs directly from public information, heralding a new paradigm in PoC generation. The dual-use implications of this emerging capability demand careful deliberation by the security community.

\clearpage
{\appendices
\section{Case Studies} \label{statistics}
As discussed in~\Cref{sec:step1}, valid PoCs emerge when CVE descriptions disclose explicit root causes or exploitation details.~\Cref{fig:RQ1_description.pdf} presents a representative CWE-89 vulnerability to illustrate this pattern.

\begin{figure}[H]
  \centering
  \includegraphics[width= 0.8\linewidth]{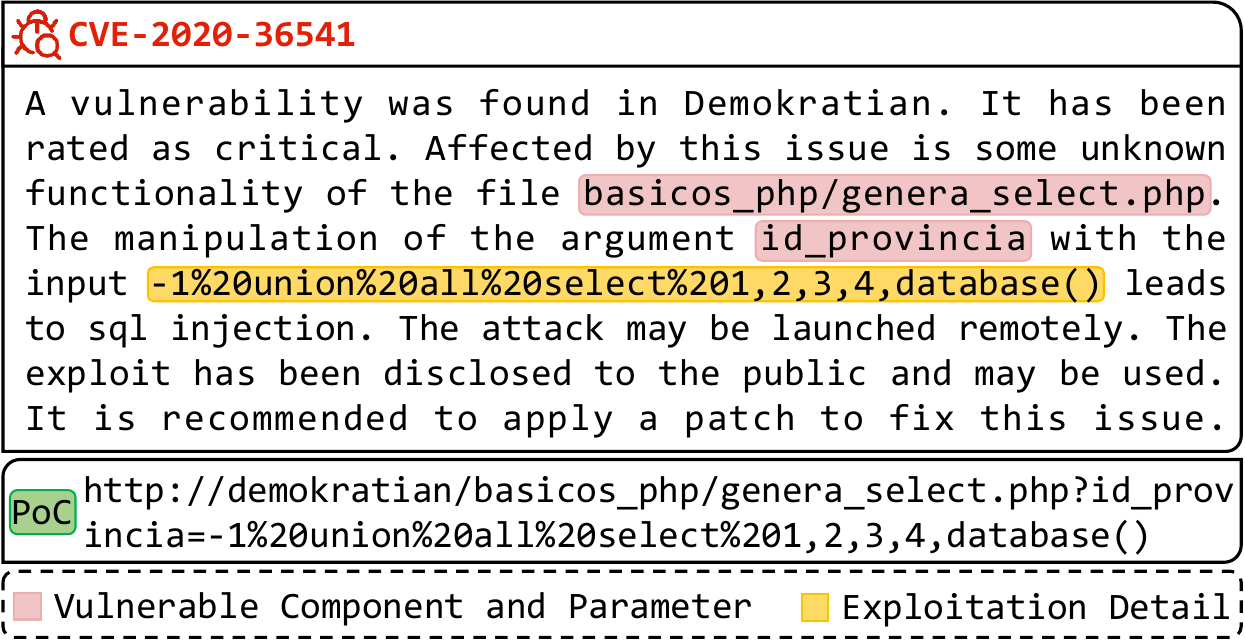}
  \caption{Case study: description-only exploitation by LLMs.}
  \label{fig:RQ1_description.pdf}
\end{figure}

As a supplement to~\Cref{sec:step1},~\Cref{fig:RQ1_Payload_R1.pdf} presents the distribution of attack payloads generated by DeepSeek-R1, highlighting its higher diversity relative to GPT-4o.

\begin{figure}[H]
  \centering
  \includegraphics[width= 0.85\linewidth]{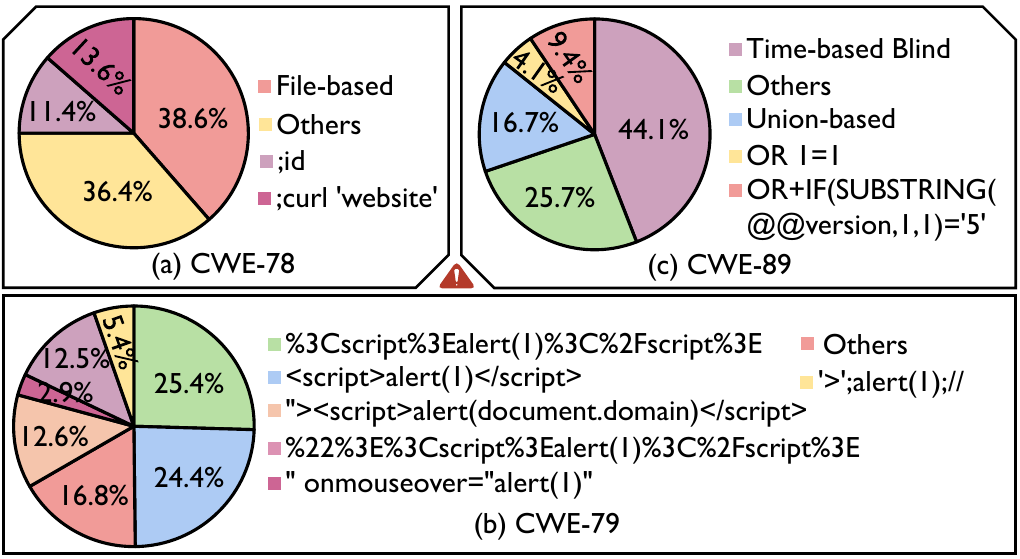}
  \caption{Distribution of taint-style vulnerability attack payloads generated by DeepSeek-R1.}
  \label{fig:RQ1_Payload_R1.pdf}
\end{figure}

% As discussed in~\Cref{sec:step1},
Additionally, \Cref{fig:RQ1_PoC_Format.pdf} briefly presents two types of PoC formats. Interaction-based PoCs mimic structured HTTP requests, while Code-style PoCs resemble directly executable scripts (e.g., HTML, Python).
\begin{figure}[H]
  \centering
  \includegraphics[width= 0.75\linewidth]{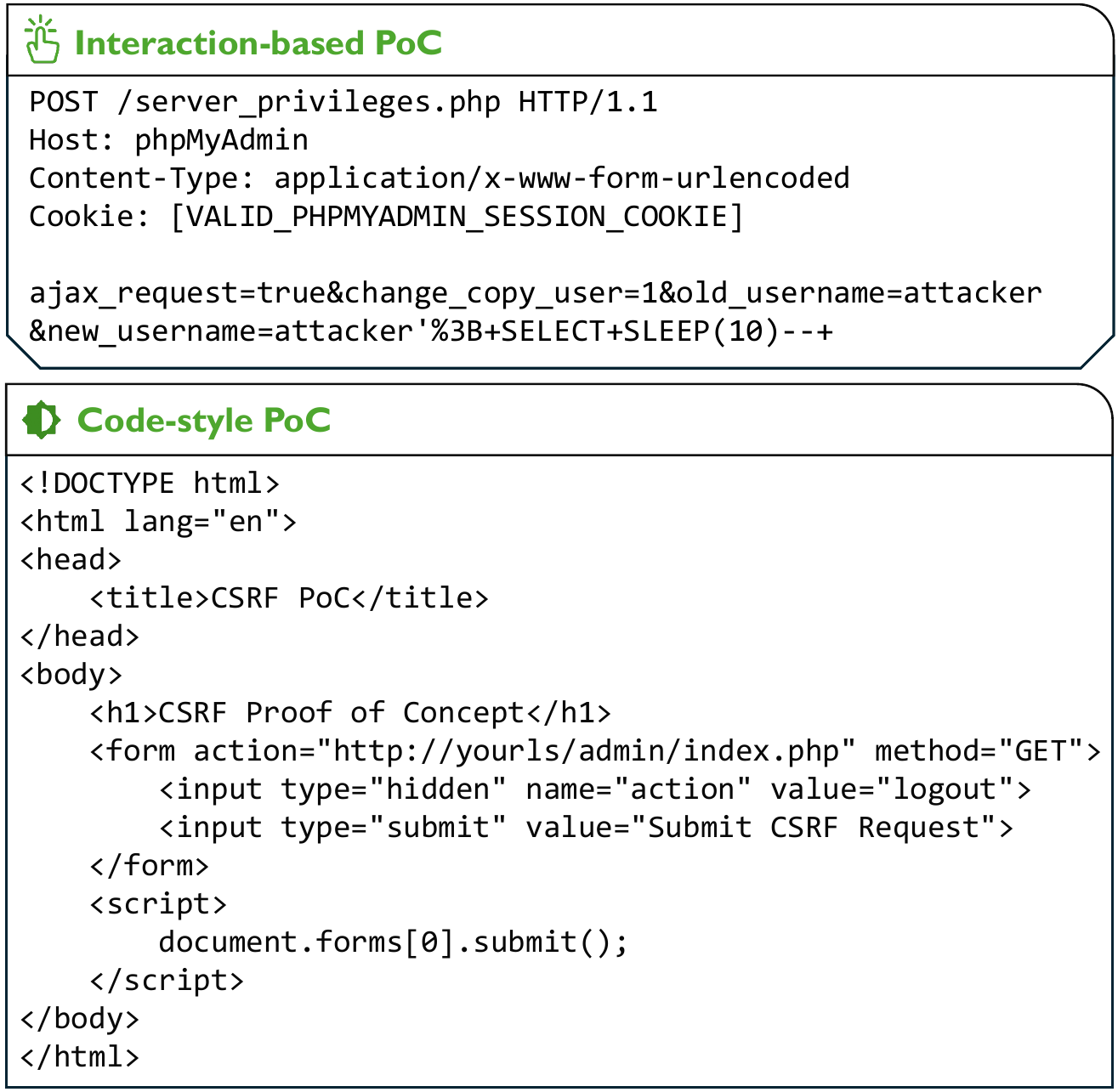}
  \caption{Examples of Interaction-based and Code-style PoC formats.}
  \label{fig:RQ1_PoC_Format.pdf}
\end{figure}

As discussed in~\Cref{sec:step2},~\Cref{fig:RQ2_Example.pdf} contrasts two CWE-79 vulnerabilities with different exploitation complexity. CVE-2017-6537 imposes multiple constraints, particularly terminating the \textless \textit{style}\textgreater tag, often leading to payload failure. In contrast, CVE-2017-6478 involves no such structural bypass, enabling straightforward exploitation.
\begin{figure}[H]
  \centering
  \includegraphics[width=0.95 \linewidth]{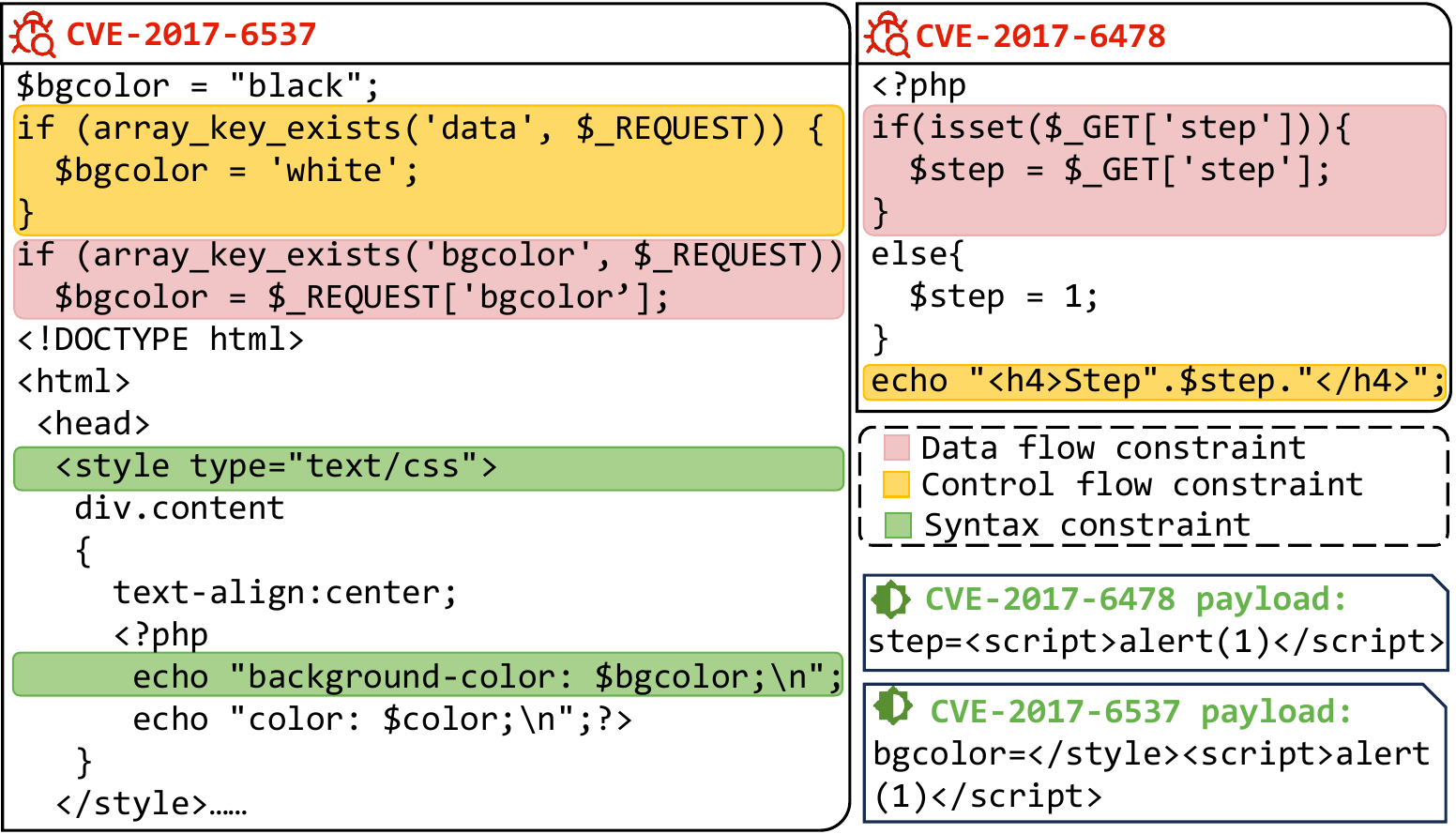}
  \caption{Case study: contrasting simple and complex exploit constraints.}
  \label{fig:RQ2_Example.pdf}
\end{figure}

\section{Prompt Templates} \label{prompts}
As a supplement to~\Cref{sec:step2}, \Cref{fig:RQ2_Prompt.pdf} presents a prompt template for extracting sub-task results in taint-style vulnerabilities using public information.
Moreover, \Cref{fig:RQ3_Prompt.pdf} presents a prompt used to directly generate PoCs when sufficient information is available (\Cref{sec:step3}).
\begin{figure}[H]
  \centering
  \includegraphics[width=0.8\linewidth]{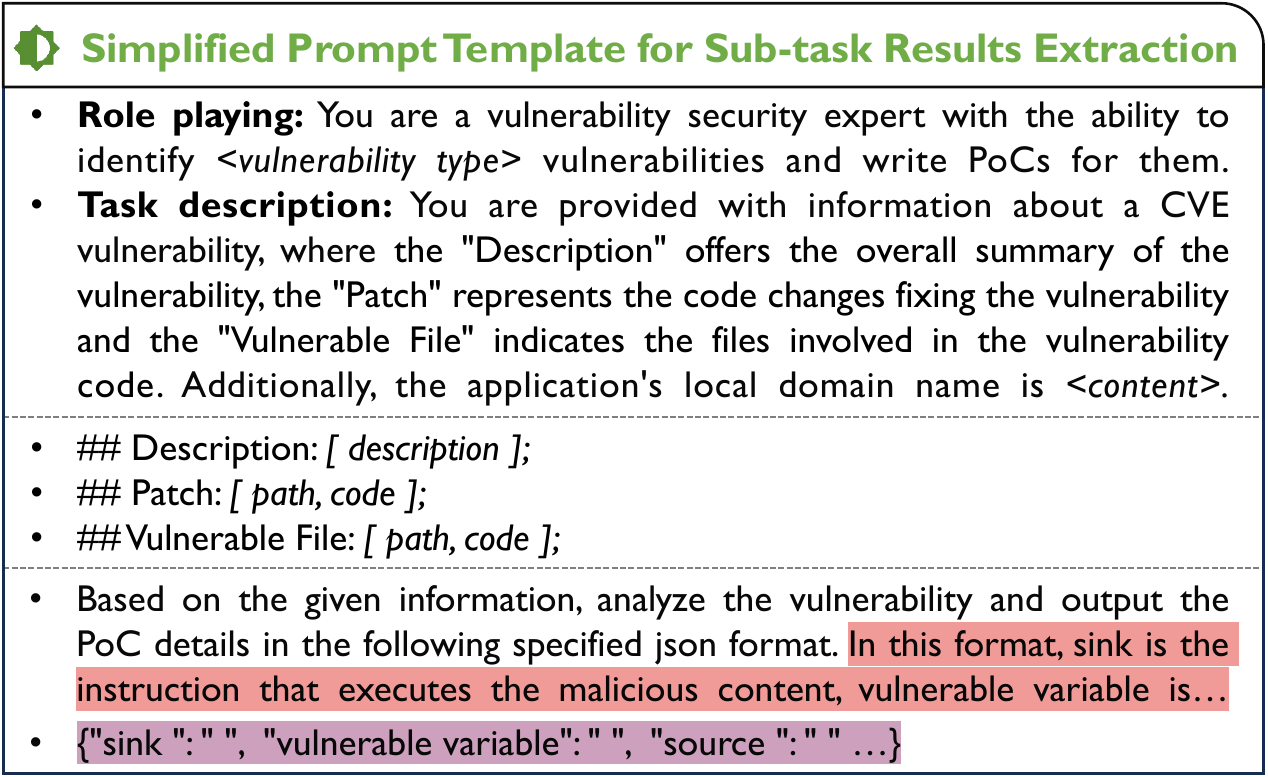}
  \caption{A simplified prompt for taint-style vulnerability sub-task extraction.}
  \label{fig:RQ2_Prompt.pdf}
\end{figure}
\begin{figure}[H]
  \centering
  \includegraphics[width=0.8\linewidth]{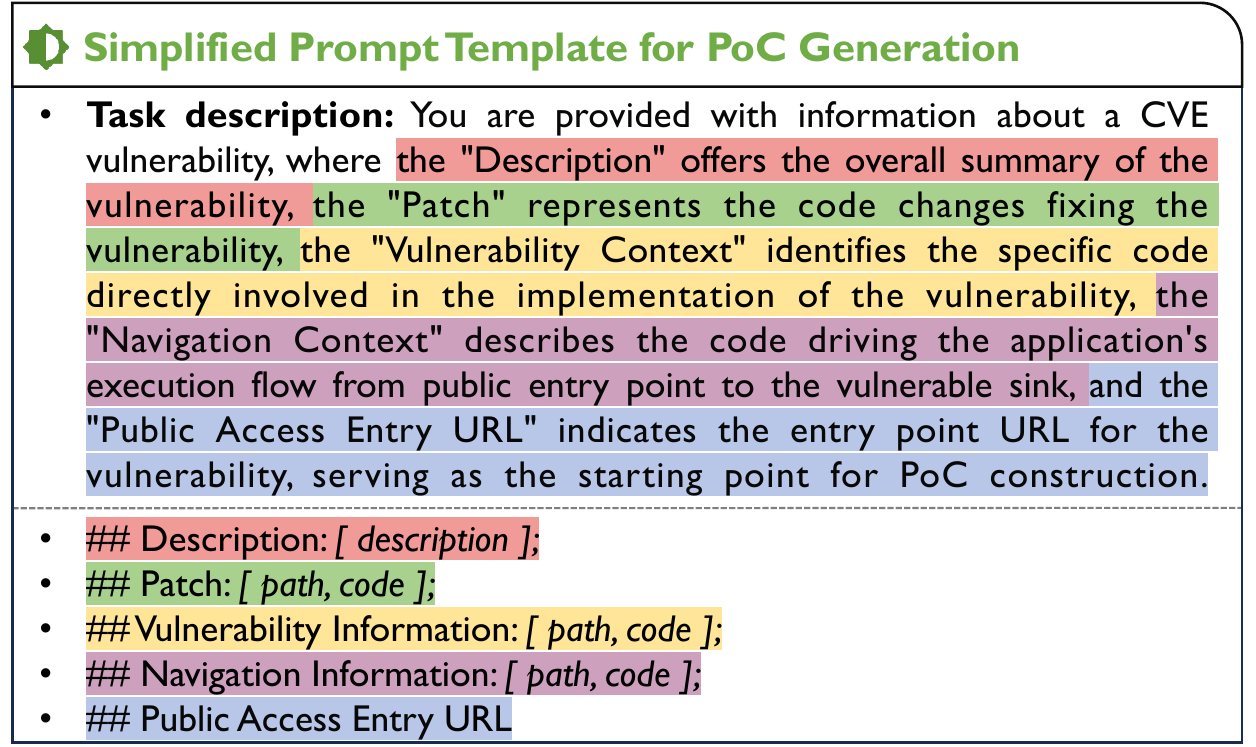}
  \caption{A simplified prompt for directly PoC generation under sufficient information.}
  % \Description{}
  \label{fig:RQ3_Prompt.pdf}
\end{figure}

% \section{Vulnerability Examples } \label{examples}
% Here are examples.
\begin{figure*}[]
  \centering
  \includegraphics[width=0.9\linewidth]{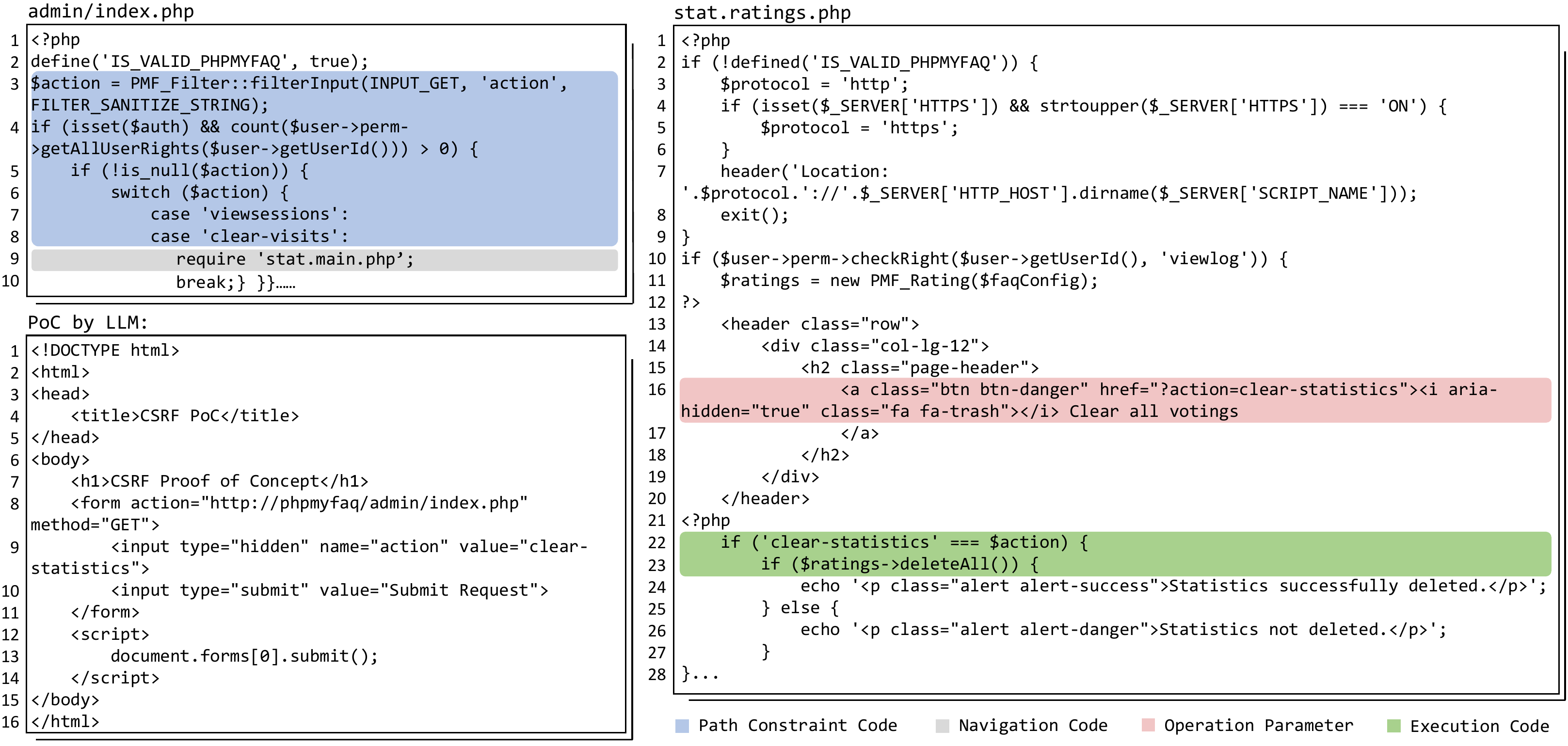}
  \caption{A simplified example for CVE-2017-15730 (CWE-352).}
  \label{fig:example_csrf.pdf}
\end{figure*}

\begin{figure*}[]
  \centering
  \includegraphics[width=0.9\linewidth]{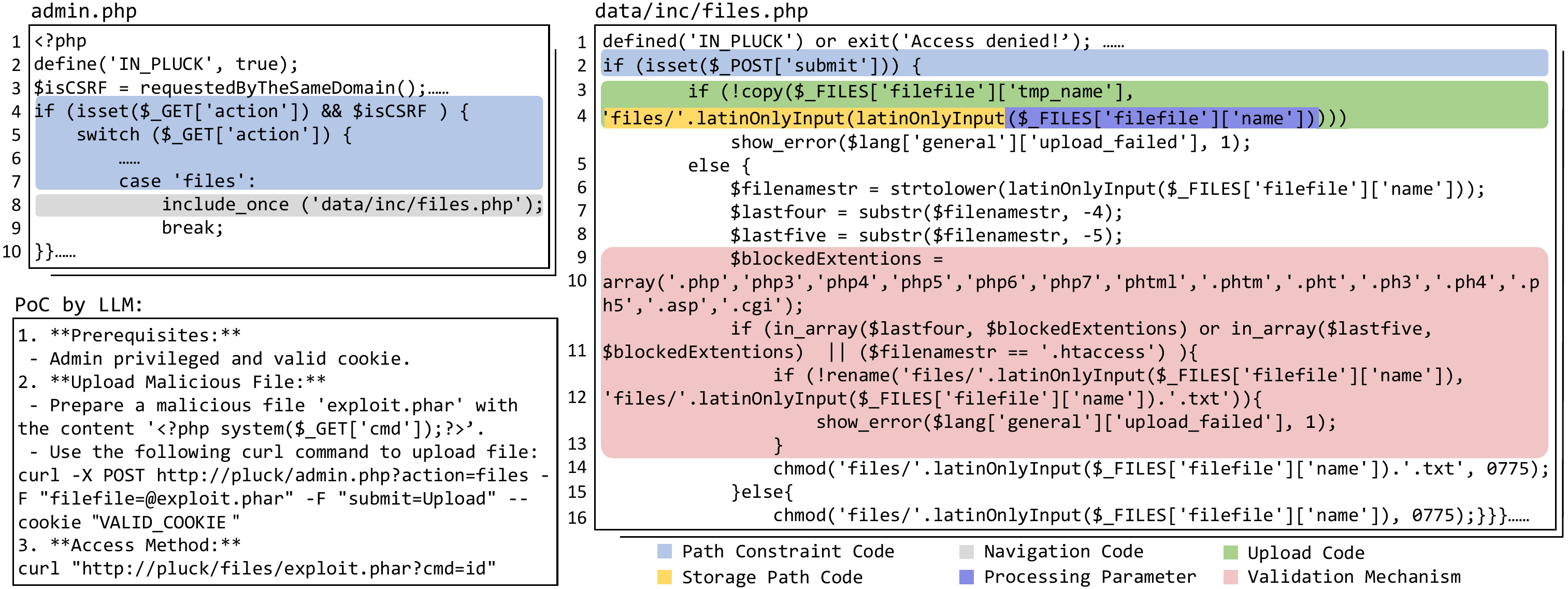}
  \caption{A simplified example for CVE-2020-29607 (CWE-434).}
  \label{fig:example_ufu.pdf}
\end{figure*}

\section{Sub-Task Decomposition} \label{Sub-task}
This appendix presents the sub-task decomposition process for PoC generation on CWE-352 and CWE-434, as part of the~\Cref{sec:step2}.

For CWE-352 vulnerabilities, we consider the traditional CSRF attack, where the attacker makes a victim’s web browser silently send a forged HTTP request to a vulnerable website and cause an undesired state-changing action~\cite{sudhodanan2017large,pellegrino2017deemon,calzavara2019mitch}. We decompose the PoC generation process into 11 sub-tasks, grouped into three modules: the attack vector crafting phase (4 tasks), the navigation path generation phase (4 tasks), and the PoC assembly phase (3 tasks). Since the navigation path generation and PoC assembly phases are consistent with those described in~\Cref{sec:step2}, we focus here on the attack vector crafting phase.

\noindent\textbf{Attack Vector Crafting Phase.} It comprises 4 sub-tasks: identifying \textit{protection mechanisms}, \textit{execution code}, \textit{operation parameters}, and reasoning about \textit{bypass techniques}. Specifically, LLMs need to first analyze the protection mechanisms before patch (e.g., \texttt{Referer/Origin header verification} or \texttt{SameSite cookies}) and reason possible bypass techniques. These techniques enable the attacker to circumvent CSRF defenses and execute unauthorized actions. Then, LLMs need to analyze the target vulnerable functionality, primarily to identify the execution code and operation parameters. As shown in~\Cref{fig:example_csrf.pdf}, it briefly shows the PoC generation process for CVE-2017-15730. Execution code snippet refers to the application logic responsible for processing the forged request, including code that handles incoming user inputs and the statements that change the application's state (e.g., \texttt{stat.ratings.php}, Line 22-23). The operation parameters define the elements required in the forged request that must be included to successfully invoke the vulnerable functionality (e.g., \texttt{action=clear-statistics in stat.ratings.php}, Line 16). The LLM is expected to identify these elements and, together with the input conditions derived from the navigation path generation phase, to construct a PoC that simulates the CSRF attack, such as by creating an auto-submitting form to exploit the identified vulnerabilities, as shown in~\Cref{fig:example_csrf.pdf}.

For CWE-434 vulnerabilities, we decompose the PoC generation process into 17 sub-tasks, organized into three modules: the attack vector crafting phase (10 tasks), the navigation path generation phase (4 tasks), and the PoC assembly phase (3 tasks). Here, we focus on the attack vector crafting phase.

\noindent\textbf{Attack Vector Crafting Phase.} It covers three core steps: constructing the malicious file (5 tasks), parsing file upload handling (3 tasks), and determining access methods (2 tasks). Constructing the malicious file involves identifying the \textit{validation mechanism}, reasoning \textit{bypass techniques}, and specifying the malicious \textit{file name}, \textit{file content-type}, and \textit{file content}. As shown in~\Cref{fig:example_ufu.pdf}, it briefly shows the PoC generation process for CVE-2020-29607. LLMs need to first analyze the application's file validation mechanism (e.g., blacklist restrictions in \texttt{data/inc/files.php}, Lines 9–13) to identify potential flaws and reason about possible bypass techniques. Then, the LLM constructs a malicious file, specifying details such as file name, file content-type, and file content to evade the identified validation checks~\cite{lee2020fuse}. Parsing file upload handling requires identifying \textit{upload code} (eg., \texttt{data/inc/files.php}, Line 3), \textit{processing parameters} (eg., \texttt{data/inc/files.php}, Line 4) and \textit{the storage path code}. Storage path code is used to compute where the uploaded file is ultimately saved (e.g., \texttt{data/inc/files.php}, Line 4). 
Finally, after combining the information from the navigation path generation and the PoC assembly phases, the LLM needs to determine how to access the uploaded file upon a successful upload. This involves two steps: reasoning its final \textit{storage location} on the server and determining the appropriate \textit{access method}.}

\bibliographystyle{IEEEtran}
\bibliography{ref}
% \end{thebibliography}
\end{document}